\documentclass{article}

\usepackage[utf8]{inputenc}
\usepackage[T1]{fontenc}

\usepackage[margin=1in]{geometry}
\usepackage{setspace}
\usepackage{multicol}

\usepackage{amsmath}
\usepackage{amssymb}
\usepackage{siunitx}

\usepackage{graphicx}
\graphicspath{{./figures/}}
\usepackage{caption}
\usepackage{subcaption}
\usepackage[dvipsnames]{xcolor}
\usepackage{rotating}
\usepackage{tikz}
\usepackage{booktabs}
\usetikzlibrary{shapes.geometric, arrows.meta, positioning}

\usepackage[numbers, square, sort&compress]{natbib}
\numberwithin{equation}{section}
\usepackage{authblk}
\usepackage{enumitem}
\usepackage{csquotes}
\usepackage{lineno}
\usepackage{totcount}
\usepackage{etoolbox}
\usepackage[toc,page]{appendix}

\definecolor{DarkBurntOrange}{RGB}{200,80,0}
\usepackage{hyperref}
\hypersetup{
  colorlinks=true,
  urlcolor=ForestGreen,
  linkcolor=DarkBurntOrange,
  citecolor=blue
}

\title{\vspace{-2cm} \textbf{Modifications of CMB Temperature and Polarization Quadrupole Signals in Thurston Spacetimes}}

\author[1]{Tanay Gupta \thanks{tanay23@iiserb.ac.in}}
\author[1]{Rajib Saha \thanks{rajib@iiserb.ac.in}}
\author[1]{Sukanta Panda \thanks{sukanta@iiserb.ac.in}}

\affil[1]{Department of Physics, Indian Institute of Science Education \& Research, Bhopal, India}

\date{}

\newcommand{\crpartial}{\textup{\rmfamily\dh}}

\newtotcounter{citcount}

\AtBeginEnvironment{thebibliography}{
  \let\oldbibitem\bibitem
  \renewcommand{\bibitem}{\stepcounter{citcount}\oldbibitem}
}

\begin{document}
\maketitle

%+++++++++++++++++++++++++++++++++++++++++++++++++++++++++++++++++++++++++++++++++++++++++++++++++++++++++++++++++++++++++++++++++++++++++++++++++++++++++++++++++++++++++++++
\vspace*{-0.8cm}
\begin{abstract}
Recent cosmological tests have discovered persistent CMB anomalies in the large-scale isotropy of the universe. Motivated thus by such anomalies that hint at potential departures from statistical isotropy present in recent cosmological observations with the advent of the `precision cosmology' era, we are led to explore the viability of anisotropic Thurston geometries, described in William Thurston’s geometrization conjecture. In this work, we examine the coherent temperature and polarization signals generated in the CMB sky by such geometries. We begin with introducing Thurston spacetimes as our background models and the formalism we use to obtain the patterns. We then construct a set of transfer equations relative to a given background and solve them for each spacetime geometry. We finally discuss the role of spatial curvature in these FLRW limiting models along with their underlying geometry, and attempt to establish some general results on the symmetries of the patterns produced by their time evolution in terms of temperature and Stokes parameters P, Q and U. We show the evolution of temperature and polarization amplitudes in terms of such Stokes parameters at different redshifts and attempt to isolate individual Thurston geometries.
\end{abstract} 

%+++++++++++++++++++++++++++++++++++++++++++++++++++++++++++++++++++++++++++++++++++++++++++++++++++++++++++++++++++++++++++++++++++++++++++++++++++++++++++++++++++++++++++++
\section{Introduction} \label{one}
The standard model of cosmology, the `$\Lambda$CDM' model, has so far been extremely successful in determining the cosmology, structure formation, and expansion of the universe, along with predicting the morphology of CMB patterns observed in the sky today \cite{adame2025desi, aghanim2020planck, abbott2019first, alam2017clustering}. The standard metric under this model, the FLRW metric, dictates a universe which is homogeneous and isotropic on large scales ($\gtrsim$ 100 - 300 Mpc \cite{einasto2025fractal, nadathur2013seeing}). However, as observational techniques have improved over the past decades, it has become clear that our cosmos is far from simple due to various anomalies. There thus quickly arose a need to develop more general models, with FLRW as a limiting case, capable of reproducing standard predictions and providing a satisfactory explanation of such anomalies not yet accounted for by our standard model. Among such models, we consider the possibility of \textit{Thurston geometries}.

Thurston geometries \cite{awwad2024large, thurston1982three} are a class of homogeneous but anisotropic spacetime models from the geometrization conjecture proposed first by W. Thurston in his work \cite{thurston1982three} and later proved by G. Perelman \cite{perelman2002entropy, perelman2003finite, perelman2003ricci}. These are a class of eight homogeneous, 3D Riemannian geometries, consisting of three isotropic and five anisotropic ones. Motivated by the shear, curvature, eccentricity and preferred-axis constraints we obtained \cite{gupta2026thurston} for \emph{truly anisotropic} Thurston geometries \cite[Sec. 6]{awwad2024large} from Pantheon+ SH0ES type Ia supernova data, we ask what coherent large-angle CMB temperature and polarization signature each geometry would produce, extending the coherent-radiation formalism of \cite{sung2011temperature} beyond the classical Bianchi types to the Thurston geometries.

We thus isolate and track the pure geometric propagation effects by evolving all the anisotropic Thurston geometries for the large-scale mode $\ell = 2$, following \cite{sung2011temperature}, for two related reasons. First, in a spatially homogeneous but anisotropic cosmology, the quadrupole is the lowest-order coherent anisotropy the geometry can source: the dipole is purely kinematic (observer motion), and higher multipoles arise only from repeated coupling in the transfer equations or a non-quadrupole seed. Second, as mentioned, our geometries are constrained by independent Pantheon+/SH0ES fits to have a radius of curvature far exceeding the Hubble radius today ($L_0 \gg R_0 \equiv c / H_0$, i.e. $|\Omega_k|\ll1$); since every curvature-dependent term in the transfer equations \cite{sung2011temperature} scales with $\sqrt{|\Omega_k|}$ or a positive power thereof, the leading anisotropic correction to an otherwise-FRW sky is precisely the $\ell=2$ term, with $\ell\geq3$ content suppressed to higher order in $\sqrt{|\Omega_k|}$. This makes our truncation consistent with, and motivated by, the same small-curvature regime these geometries are independently constrained to occupy. We form and solve the transfer (Boltzmann) equations using the model-independent methodology described in \cite{sung2011temperature}. The resulting maps illustrate how spatial anisotropy distorts radiation fields over cosmic time, laying the necessary framework for future full-sky statistical and power-spectrum comparisons.

The paper is organized as follows: in section \ref{two}, we introduce Thurston geometries; in section \ref{three}, we present the basic formalism we used to derive all the equations necessary for each geometry, derive all our evolutions and discuss the various scattering as well as initial seeding mechanisms; and in section \ref{four}, we present the results of the temperature and polarization sky maps in each of the Thurston universes. Finally, in section \ref{five}, we discuss our observations and conclude this work. Throughout this paper, Greek indices ($\mu, \nu, \alpha, \beta = 0, 1, 2, 3$) denote 4D coordinate frame components, Latin indices ($a, b, c, d = 0, 1, 2, 3$) denote 4D local tetrad frame components, and lower-case spatial indices ($i, j, k = 1, 2, 3$) denote 3D spatial components. Also, natural units are employed ($8\pi G = c = k_b = \hbar = 1$).

%+++++++++++++++++++++++++++++++++++++++++++++++++++++++++++++++++++++++++++++++++++++++++++++++++++++++++++++++++++++++++++++++++++++++++++++++++++++++++++++++++++++++++++++
\section{Thurston-Perelman's geometrization theorem} \label{two}
The Thurston-Perelman geometrization theorem \cite{awwad2024large} is a partial classification of three-dimensional manifolds, analogous
to the uniformization theorem that classifies the possible geometries of the Riemannian surfaces. The main difference lies in the fact that not every 3-manifold can be endowed with a unique geometry, but rather every 3-manifold can be cut into pieces that can be endowed with one.

Any maximal, simply connected, three-dimensional geometry X that admits a compact quotient is equivalent to one of the eight geometries below:
\begin{multicols}{4}
\begin{enumerate}
    \item $\mathbb{R}^3$
    \item $S^3$
    \item $\mathbb{H}^3$
    \item $\mathbb{R} \times \mathbb{H}^2$
    \item $\mathbb{R} \times S^2$
    \item $\widetilde{U \left(\mathbb{H}^2 \right)}$
    \item Nil
    \item Solv
\end{enumerate}
\end{multicols}
These eight maximal geometries can be said to form the building blocks of all compact 3-manifolds and are referred to as \textit{Thurston geometries}. These are:
\begin{enumerate}
    \item FLRW spacetimes $\left(\mathbb{R}^3/ \mathbb{H}^3/ S^3\right)$ \hspace{0.02cm} \textbf{(3)}
        \begin{equation} \label{flrw}
            ds^2 = -dt^2 + a^2 (t) \{d\chi^2 + S_\kappa^2(\chi) d\Omega^2\}
        \end{equation}
    \item 2D maximally symmetric spatial sectors combined with an additional, orthogonal flat line $\left(\mathbb{R} \times \mathbb{H}^2/S^2\right)$ \hspace{0.02cm} \textbf{(2)}
        \begin{equation} \label{rh2s2}
            ds^2 = -dt^2 + a^2 (t) \{dz^2 + d\chi^2 + S_\kappa^2(\chi) d\phi^2\}
        \end{equation}
        \{z $\in$ R is orthogonal to ($\chi$,$\phi$) plane\}\\  
        where $\kappa$ is the curvature parameter of the universe and is related to the radius of curvature of each Thurston geometry L by \cite{awwad2024large}
        \begin{equation} \label{kL}
            L = \frac{1}{\sqrt{\left|k \right|}}
        \end{equation}
        where
        \begin{equation}
        	\kappa =
        	\begin{cases}
        		< 0, & \mathbb{H}^3, \mathbb{R} \times \mathbb{H}^2, \widetilde{U \left(\mathbb{H}^2 \right)}, \text{Nil}, \text{Solv} \\
        		0, & \mathbb{R}^3 \\
        		> 0, & S^3, \mathbb{R} \times S^2
        	\end{cases}
        \end{equation}
    \item Universal cover of the unit tangent bundle of the hyperbolic plane $\left(\widetilde{U(\mathbb{H}^2)}\right)$ \hspace{0.02cm} \textbf{(1)}
        \begin{equation} \label{uh2}
            ds^2 = -dt^2 + a^2(t)\ \left\{dx^2 + \cosh^2 \left(x\sqrt{-\kappa}\right) dy^2 + \left(dz + \sinh\left(x\sqrt{-\kappa}\right)dy \right)^2\ \right\}
        \end{equation}
    \item Nilpotent subgroup of an extension of the group of isometries (abb. \textit{Nil}) \hspace{0.02cm} \textbf{(1)}
        \begin{equation} \label{nil}
            ds^2 = -dt^2 + a^2 (t) \left\{dx^2 + \left(1 - \kappa \, x^2 \right)dy^2 + dz^2 - 2 \, x \, \sqrt{-\kappa} \, dy \, dz \right\}
        \end{equation}
    \item Solvable Lie group (abb. \textit{Solv}) \hspace{0.02cm} \textbf{(1)}
        \begin{equation} \label{solv}
            ds^2 = -dt^2 + a^2 (t) \{e^{2z\sqrt{-\kappa}}dx^2 + e^{-2z\sqrt{-\kappa}}dy^2 + dz^2\}
        \end{equation}
        where
        \begin{equation}
        	S_\kappa (\chi) =
        	\begin{cases}
        		\frac{\sin \left(\chi \sqrt{\kappa} \right)}{\sqrt{\kappa}}, & \kappa > 0 \hspace{2mm} (S^3, \hspace{1mm} \mathbb{R} \times S^2)\\
        		\chi, & \kappa = 0\\
        		\frac{\sinh \left(\chi \sqrt{-\kappa} \right)}{\sqrt{-\kappa}}, & \kappa < 0 \hspace{2mm} (\mathbb{H}^3, \hspace{1mm} \mathbb{R} \times \mathbb{H}^2)
        	\end{cases} 
        \end{equation} 
\end{enumerate}
In the above geometries, the Nil geometry is modeled on the three-dimensional Heisenberg group (a 2-step nilpotent Lie group); it can be viewed as a non-trivial real line bundle over the Euclidean plane $\mathbb{R}^2$. Geometrically, it exhibits translation invariance in the plane alongside a non-commutative central shear along the fiber axis. The Solv geometry is modeled on the solvable Lie group $E(1,1)$ (the group of rigid motions of 2D Minkowski space), Solv possesses a semidirect product structure $\mathbb{R}^2 \rtimes \mathbb{R}$. Geometrically, it is characterized by strong spatial anisotropy: as one moves along the $z$-axis, orthogonal spatial directions expand and contract exponentially as $e^z$ and $e^{-z}$, respectively.

The most general fluid solution compatible with the above geometries is given by \cite{awwad2024large}
\begin{equation}
	T^\mu _\nu = \rho \, u^\mu \, u_\nu + p \, \delta^\mu _\nu + \pi^\mu _\nu
\end{equation}
where $\rho = \rho$(t), p = p(t) denote the fluid energy density and pressure, respectively; $u^\mu$ = $u^\mu$(t) is the velocity vector of the fluid and $\pi^\mu _\nu$ is the shear tensor, which is symmetric $\left(\pi_{\mu\nu} = \pi_{\nu\mu} \right)$, traceless $\left(\pi^\mu _\mu = 0 \right)$ and transverse $\left(\pi^\mu _\nu \, u^\nu = 0 \right)$.

We thus distinguish the three-dimensional spatial geometries $\Sigma^3 \in \{\mathbb{R}^3, \mathbb{H}^3, \mathbb{S}^3, \mathbb{R} \times \mathbb{H}^2, \mathbb{R} \times \mathbb{S}^2, \widetilde{U \left(\mathbb{H}^2 \right)}, \mathrm{Nil}, \mathrm{Solv}\}$ from the corresponding four-dimensional cosmological spacetimes $M^4 = I \times \Sigma^3$ governed by line elements of the form $ds^2 = -dt^2 + a^2(t) d\sigma^2_{\Sigma^3}$. As is evident, all the anisotropic Thurston geometries \eqref{rh2s2}, \eqref{uh2} - \eqref{solv} reduce to flat FLRW in the limit $\kappa \to$ 0.

%+++++++++++++++++++++++++++++++++++++++++++++++++++++++++++++++++++++++++++++++++++++++++++++++++++++++++++++++++++++++++++++++++++++++++++++++++++++++++++++++++++++++++++++
\section{Formalism} \label{three}
To analyze photon geodesics in each of our geometries, we introduce a tetrad basis constructed from a local coordinate system $x^\alpha$ \cite{ellis1969class, sung2011temperature} by
\begin{equation} \label{tetrad}
    \mathbf{e}_a = \mathrm{e}^\alpha _a \frac{\partial}{\partial x^\alpha}
\end{equation}
such that
\begin{equation}
    g_{\mu\nu} = e^a_\mu e^b_\nu \eta_{ab} = \text{diag}(-1, +1, +1, +1)
\end{equation}
meaning that the tetrad basis $\mathbf{e}_a$ is orthonormal. The Ricci rotation coefficients (not to be confused with II order Christoffel symbols, used later, containing one index raised) are defined by
\begin{equation}
    \boxed{\Gamma_{abc} = e_{a \mu ; \nu} \, e^\mu _b \, e^\nu _c}
\end{equation}
in which semicolons denote covariant derivatives. We now proceed for Thurston geometries \eqref{flrw}-\eqref{solv} as demonstrated in the following sections. 

Our general methodology will be to generalize the Liouville equation to provide a complete description of photons traveling through curved spacetime. After that, we will solve our equations in the tetrad frame \eqref{tetrad}. This requires us to set up radiation distribution functions that incorporate all Stokes parameters needed to specify polarized radiation. We also need to include a source term that accounts for Thomson scattering by free electrons over the entire history from decoupling to the present epoch.

%+++++++++++++++++++++++++++++++++++++++++++++++++++++++++++++++++++++++++++++++++++++++++++++++++++++++++++++++++++++++++++++++++++++++++++++++++++++++++++++++++++++++++++++
\subsection{Radiation description}
The linear perturbation of the photon phase-space distribution is characterized by the four standard Stokes parameters $\mathbf{S} = (I, Q, U, V)^T$, which represent intensity differences in a local orthonormal basis $(\mathbf{e}_1, \mathbf{e}_2)$ orthogonal to the photon propagation vector $\hat{\mathbf{n}}$:
\begin{enumerate}
	\item $I = I_x + I_y$: Total radiation intensity (temperature perturbation $\Delta T/T_0$).
	\item $Q = I_x - I_y$: Intensity difference between linear polarization along $\mathbf{e}_1$ ($0^\circ$) and $\mathbf{e}_2$ ($90^\circ$).
	\item $U = I_{+45^\circ} - I_{-45^\circ}$: Intensity difference between linear polarization along axes rotated by $\pm 45^\circ$.
	\item $V = I_{\mathrm{R}} - I_{\mathrm{L}}$: Intensity difference between right- and left-circularly polarized components.
\end{enumerate}
and on spin-weighted spherical harmonics. The transfer equation for polarized radiation propagating through spacetime can be described by a (complex) photon distribution comprising components with dimensionless spin weights 0 (unpolarized) and 2 (polarized), i.e.
\begin{equation} \label{NV}
    \hat{N} \equiv
    \begin{pmatrix}
    N^0 \\
    N^2
    \end{pmatrix}
    = \frac{c^2}{2 h \nu^3}
    \begin{pmatrix}
    I + iV \\
    Q - iU
    \end{pmatrix}
\end{equation}
where `$\nu$' represents the physical photon frequency at the instantaneous (local) epoch where the radiation is evaluated \cite{dautcourt1978polarized}, and here we consider it to be the CMB detector frequency $\left(\approx 150 \, \text{GHz} \right)$, `T' is the local temperature and rest are usual physical constants. $\hat{N}$ is dimensionless as relative temperature fluctuations (see \S\ref{form} and our initial seeds code in \S \ref{code}), while Stokes parameters I, Q, U \& V are in the absolute intensity units $\left(\text{W} \, \text{sr}^{-1} \, \text{m}^{-2} \, \text{Hz}^{-1} \right)$, while the photon multipoles $\hat{N}$ dimensionless (see equations \eqref{one_to_one} in section \ref{form}).

We describe the amplitude of linear polarization P and the polarization orientation angle $\chi$ as
\begin{equation}
    P = \frac{\sqrt{Q^2 + U^2}}{I}
\end{equation}
\begin{equation}
    \chi = \frac{1}{2} \, \text{arctan} \, \frac{U}{Q}
\end{equation}
The total change of photon distribution function $N = N \left( x^\mu, p^\mu \right)$ is given by
\begin{equation} \label{main}
\begin{split}
    \frac{d \hat{N}}{d\lambda} &= \frac{\partial \hat{N}}{\partial x^\mu}\frac{d x^\mu}{d \lambda} + \frac{\partial N}{\partial p^\alpha}\frac{dp^\alpha}{d\lambda}\\
    &= \frac{\partial \hat{N}}{\partial x^\mu}\frac{d x^\mu}{d x^\alpha}\frac{d x^\alpha}{d\lambda} + \frac{\partial \hat{N}}{\partial p^\alpha}\frac{dp^\alpha}{d\lambda}\\
    &= l^\mu _{(\alpha)}p^\alpha \frac{\partial \hat{N}}{\partial x^\mu} + \frac{\partial \hat{N}}{\partial p^\alpha}\frac{dp^\alpha}{d\lambda}\\
\end{split}
\end{equation}
where $l^a$ are spacetime unit vectors, $l^\mu _{\left(\alpha \right)} = dx^\mu / dx^\alpha$. We define the components of four-momentum as
\begin{equation} \label{p}
    \left\{p^\alpha \right\} = (\epsilon, \epsilon \hat{k}^i)
\end{equation}
where $\epsilon$ is the energy of photon and
\begin{equation} \label{ray}
	\hat{k}^i = (\sin \theta \cos \phi, \sin \theta \sin \phi, \cos \theta)
\end{equation}
is the photon ray direction. 

Using the geodesic equation for photon
\begin{equation} \label{geoeqn}
	\frac{dp^\alpha}{d\lambda} = -\Gamma^\alpha _{\beta\gamma} \, p^\beta \, p^\gamma
\end{equation}
where $p^\alpha \equiv dx^\alpha / d\lambda$ represents the components of photon four-momentum, equation \eqref{main} can be written as
\begin{equation}
	\frac{d \hat{N}}{d\lambda} = p^\alpha l^\mu _{(\alpha)} \frac{\partial \hat{N}}{\partial x^\mu} - \Gamma^\alpha_{\beta \gamma} \, p^\beta \, p^\gamma \, \frac{\partial \hat{N}}{\partial p^\alpha}
\end{equation}
If the change in N along a photon path arises from collisions only, one obtains the following Boltzmann equation:
\begin{equation} \label{boltz}
	\boxed{\frac{1}{\epsilon}\frac{d \hat{N}}{d\lambda} = l^\rho _{(0)} \frac{\partial \hat{N}}{\partial x^\rho} + \hat{k}^i l^\rho _{(i)}\frac{\partial \hat{N}}{\partial x^\rho} - \epsilon \Bar{\gamma}^0 \frac{\partial \hat{N}}{\partial \epsilon} + \frac{\Bar{\gamma}^i}{\sqrt{2}} (m^i \Bar{\crpartial} + \Bar{m}^i\crpartial) \hat{N}}
\end{equation}
where
\begin{subequations}
\begin{gather}
	\Bar{\gamma}^\alpha = \Gamma^\alpha_{00} + \Gamma^\alpha_{0i}k^i + \Gamma^\alpha_{i0}k^i + \Gamma^\alpha_{ik}k^ik^k \label{gamma_def} \\
	m^i = \frac{1}{\sqrt{2}} \, \left(\frac{\partial \hat{k}^i}{\partial \theta} + \frac{i}{\sin \theta} \frac{\partial \hat{k}^i}{\partial \phi} \right) \label{mi} \\
	\crpartial = -\left(\frac{\partial}{\partial \theta} + \frac{i}{\text{sin}\theta}\frac{\partial}{\partial \phi}\right) \text{\cite{dautcourt1978polarized,newman1966note}}
\end{gather}
\end{subequations}
As mentioned earlier, for polarized radiation, we need to extend the description of the radiation field to include both spin-0 and spin-2 components. This requires us to generalize N, which can be decomposed into parts $N^0$ and $N^2$. In the transfer equation for $N^2$, thus, these give rise to additional terms related to the change of angles $\theta$ and $\phi$, and an extra rotation $\psi$ of polarization. From the relation between the spin 0 and 2 operators, the angular derivative term is given as
\begin{equation}
    \frac{\gamma^i}{\sqrt{2}}(m^i\Bar{\crpartial}_2 + \Bar{m}^i \crpartial_2)N^2 = \frac{2i}{\epsilon} \, \cos{\theta} \, \frac{d\phi}{d\lambda} \, N^2 + \frac{\gamma^i}{\sqrt{2}} \, \left(m^i \Bar{\crpartial} + \Bar{m}^i \crpartial \right) \, N^2
\end{equation}
We must replace $N^2$ by $N^2 e^{-2i\psi}$ for the relative twisting of the directions $m^i$ and a parallel-propagated direction. This adds to the Liouville equation a term such as
\begin{equation}
    -2iN^2 \frac{d\psi}{d\lambda} = -2i \, \cot{\theta} \, \frac{d\phi}{d\lambda} \, N^2 - 2N^2 \, m^i \, \Bar{m}^k \, \epsilon \, \left(\Gamma^k _{0i} + k^l \Gamma^k_{li} \right)
\end{equation}
First two extra terms cancel out in the Liouville equation, then we obtain the simplified form
\begin{equation}
    \mathcal{D}_AN^a \equiv e^\alpha _0 \, \frac{\partial N^A}{\partial x^\alpha} + \hat{k}^i \, e^\alpha _i \, \frac{\partial N^A}{\partial x^\alpha} - \epsilon \, \gamma^0 \, \frac{\partial N^A}{\partial \epsilon} + \vartheta N^A + i \, \delta^2 _A \, N^2 \, \left(\Gamma^k_{0i} \, \epsilon^{ikl} \, \hat{k}^l + \Gamma^k_{li} \, \hat{k}^l \, \hat{k}^m \, \epsilon^{ikm} \right)
\end{equation}
where
\begin{equation}
    \frac{1}{2}(m^l \Bar{m}^j - \Bar{m}^l m^j) = -\frac{i}{2} \, \epsilon^{ljk} \, k^k
\end{equation}
and
\begin{equation}
    \vartheta = \frac{\gamma^i}{\sqrt{2}} (m^i \Bar{\crpartial} + \Bar{m}^i \crpartial)
\end{equation}

%+++++++++++++++++++++++++++++++++++++++++++++++++++++++++++++++++++++++++++++++++++++++++++++++++++++++++++++++++++++++++++++++++++++++++++++++++++++++++++++++++++++++++++++
\subsection{Initial seeding} \label{form}
We begin by writing the relative temperature fluctuations for $\ell = 2$ as
\begin{equation} \label{formula}
	\begin{split}
		\frac{\Delta T}{T} \, \left(\theta, \phi \right) &= \sum^2 _{m=-2} \, a_{2,m} \, Y_{2,m} \, \left(\theta, \phi \right) \text{\cite{bennett1996four}} \\
		&= \sum^3 _{i,j=1} \, \Re \{N^0 _{ij} \} \, \left(k^i k^j - \frac{1}{3}\delta^{ij} \right) \text{\cite{demianski2007extensions}}
	\end{split}
\end{equation}
since $\Delta T / T \, \left(\theta, \phi \right) \in \Re$, and the latter of the above two formulae is used to evaluate relative temperature fluctuations $\left( \Delta T / T \, \left(\theta, \phi \right) \right)$ maps, using the numerical solutions $\Re \{N^0 _{ij} \}$. Expanding the former of the above subequations using the reality condition of spherical harmonics
\begin{subequations}
	\begin{gather}
		Y_{l,-m} = \left(-1 \right)^m \, Y^* _{l,m} \\
		a_{l,-m} = \left(-1 \right)^m \, a^* _{l,m} \label{reality_a}
	\end{gather}
\end{subequations}
while the latter using equation \eqref{ray}, symmetry condition $\left(N^0 _{ij} = N^0 _{ji} \right)$ and traceless condition $\left(\delta^{ij} N^0 _{ij} = 0 \right)$, we obtain for $\sum^2 _{m=-2} \, a_{2,m} \, Y_{2,m} \, \left(\theta, \phi \right)$ as
\begin{equation} \label{a2m}
	\frac{\Delta T}{T} \, \left(\theta, \phi \right) = 2 \Re \{a_{2,2} \} \Re \{Y_{2,2} \} - 2 \Im \{a_{2,2} \} \Im \{Y_{2,2} \} + 2 \Re \{a_{2,1} \} \Re \{Y_{2,1} \} - 2 \Im \{a_{2,1} \} \Im \{Y_{2,1} \} + a_{2,0} Y_{2,0}
\end{equation}
since $a_{2,0}$ and $Y_{2,0}$ are real, while that for $\sum^3 _{i,j=1} \, \Re \{N^0 _{ij} \} \, \left(k^i k^j - \delta^{ij}/3 \right)$ as
\begin{equation} \label{R0_ij}
	\begin{split}
		\frac{\Delta T}{T} \, \left(\theta, \phi \right) &= \frac{1}{2} \left(\Re \{N^0 _{11}\} - \Re \{N^0 _{22}\} \right) \sin^2 \theta \cos 2\phi + \frac{1}{2} \left(\Re \{N^0 _{11}\} + \Re \{N^0 _{22}\} \right) \left(1 - 3 \cos^2 \theta \right) \\
		&+ \Re \{N^0 _{12}\} \sin^2 \theta \sin 2\phi + \Re \{N^0 _{13}\} \sin 2\theta \cos \phi + \Re \{N^0 _{23}\} \sin 2\theta \sin \phi
	\end{split}
\end{equation}
Substituting for $\Re \{Y_{2,2}\}$, $\Im \{Y_{2,2}\}$, $\Re \{Y_{2,1}\}$, $\Im \{Y_{2,1}\}$ and $Y_{2,0}$, and comparing the corresponding coefficients in equations \eqref{a2m} and \eqref{R0_ij}, we finally obtain
\begin{subequations} \label{one_to_one}
	\begin{gather}
		\Re \{N^0 _{11} \} - \Re \{N^0 _{22} \} = \sqrt{\frac{15}{2 \pi}} \, \Re \{a_{2,2} \} \\
		\Re \{N^0 _{11} \} + \Re \{N^0 _{22} \} = -\frac{1}{2} \, \sqrt{\frac{5}{\pi}} \, a_{2,0} \\
		\Re \{N^0 _{12} \} = -\frac{1}{2} \, \sqrt{\frac{15}{2 \pi}} \, \Im \{a_{2,2} \} \\
		\Re \{N^0 _{13} \} = -\frac{1}{2} \, \sqrt{\frac{15}{2 \pi}} \, \Re \{a_{2,1} \} \\
		\Re \{N^0 _{23} \} = \frac{1}{2} \, \sqrt{\frac{15}{2 \pi}} \, \Im \{a_{2,1} \}
	\end{gather}
\end{subequations}
where all the quantities on both sides are real and dimensionless. All the $a_{2,m}$s are given by the square root of
\begin{equation}
	C^\text{TT} _2 = \frac{4 \pi}{25} \int^\infty _0 \frac{dk}{k} \, \mathcal{P}_\mathcal{R} (k) \, \left[j_2 \left\{k \left(\eta_0 - \eta_e \right) \right\} \right]^2
\end{equation}
where
\begin{itemize}
	\item $C^\text{TT} _2$ = Variance of quadrupole temperature anisotropy
	\item k = Fourier modes
	\item $\mathcal{P}_\mathcal{R} (k)$ = $A_s \, \left(k / k^* \right)^{n_s - 1}$, where the pivot scale $k^*$ is taken to be about $0.05 \, \text{Mpc}^{-1}$
	\item $A_s$ = Scalar amplitude, taken to be $2.1 \times 10^{-9}$
	\item $n_s$ = Spectral tilt, taken to be 0.964
	\item $j_2$ = Spherical Bessel function of order 2
	\item $\eta_0 - \eta_e$ = Comoving distance travelled by a photon before reaching an electron just before recombination, calculated as $\left(c/H_0 \right) \times \left(1 / \left. \tau \right|_{z = 1100}  \right) \times \left. \left(1 + z \right) \right|_{z = 1100} \simeq 1.5 \, \text{Mpc}$
\end{itemize}
to yield
\begin{equation}
	\boxed{a_{2,m} = 8.768 \times 10^{-6}}
\end{equation}
at recombination. Thus,
\begin{enumerate}
	\item For m = 0, only $\Re \{N^0 _{11} \} = \Re \{N^0 _{22} \} = -2.765 \times 10^{-6}$, $\Re \{N^0 _{33}\} = - \left(\Re \{N^0 _{11}\} + \Re \{N^0 _{22}\} \right)$ get seeded
	\item for m = 1, only $\Re \{N^0 _{13}\} = \Re \{N^0 _{31}\} = -6.774 \times 10^{-6}$ get seeded, and
	\item For m = 2, only $\Re \{N^0 _{11}\} = -\Re \{N^0 _{22}\} = 6.774 \times 10^{-6}$ get seeded 
\end{enumerate}
while those of $m < 0$ are automatically considered by $\mathtt{healpy}$ using the reality condition \eqref{reality_a}. Given that our chosen geometries (equations \eqref{flrw} - \eqref{solv}) incorporate an inherent geometrical, instead of kinematical, shear with an isotropic expansion \cite{awwad2024large}, we excite the monopole $N^0 _0$ also of the same order as the quadrupoles n order to get interesting higher-order patterns in the radiation background \cite{sung2011temperature}. 

%+++++++++++++++++++++++++++++++++++++++++++++++++++++++++++++++++++++++++++++++++++++++++++++++++++++++++++++++++++++++++++++++++++++++++++++++++++++++++++++++++++++++++++++
\subsection{Scattering}
We obtain the Boltzmann equation by the addition of a source term which describes the Thomson scattering by free electrons to the right-hand side of the Liouville equation, i.e.
\begin{equation}
    \mathcal{D}_A N^A = \tau \left(-N^A + J_A \right)
\end{equation}
where the differential scattering rate $\tau$ is given by
\begin{equation} \label{tau_camb}
	\tau \equiv \frac{d\tau_{\rm opt}}{dt} = n_e \, \sigma_T \, c = \frac{1}{a}\,\frac{d\tau_{\rm opt}}{d\eta},
\end{equation}
with $n_e = X_e \, n_b$ the free electron number density, $\sigma_T$ the Thomson cross-section, $\eta$ the conformal time and $\tau_{\rm opt}$ the optical depth. The quantity $d\tau_{\rm opt}/d\eta = a\, n_e \sigma_T$ (the opacity) is computed numerically with the Boltzmann code \texttt{CAMB} \cite{lewis2000efficient, howlett2012cmb}, which solves the full recombination history (hydrogen and helium) and includes reionization, with its midpoint fixed by the optical depth $\tau_{\rm reio} = 0.054$. The opacity is tabulated on a grid uniformly spaced in $\ln(1+z)$ for $0 \leq z \leq 3000$, converted to cosmic time using $c = 9.7156 \times 10^{-15}\ \text{Mpc s}^{-1}$ and the factor $1/a = 1+z$, and then interpolated in $\ln\tau$ against $\ln(1+z)$ during the numerical integration. The cosmological parameters are $\Omega_b h^2 = 0.023$, $\Omega_m = 0.315$ and $\Omega_k = 0.044$, with $h$ fixed by the adopted $H_0$.

Two other important considerations follow. Firstly, we fix the background expansion across all spatial geometries, in order to isolate the purely propagational effects of the Thurston geometries on radiative transfer. Including radiation together with geometry-dependent curvature leads to scale factors that are analytically intractable (involving elliptic functions) and to numerical instabilities. We therefore evolve the background with a matter, curvature and cosmological-constant Friedmann equation, neglecting radiation:
\begin{equation} \label{simpl1}
	H(a) \equiv \frac{\dot a}{a} = H_0 \sqrt{\Omega_M \, a^{-3} + \Omega_k \, a^{-2} + \Omega_\Lambda}\,, \qquad \Omega_\Lambda = 1 - \Omega_M - \Omega_k \,,
\end{equation}
with $\Omega_M = 0.315$, $\Omega_k = 0.044$ and $\Omega_\Lambda = 0.641$. The scale factor $a = 1/(1+z)$ is used as the independent variable, so that the time derivative is replaced by
\begin{equation}
	\frac{d}{dt} = a H(a) \, \frac{d}{da},
\end{equation}
and the comoving radial distance is obtained from $\chi(a) = \int_a^{1} da' / \left[a'^2 H(a')\right]$. For $\Omega_k \to 0$, Eq.~\eqref{simpl1} reduces to the closed-form flat solution
\begin{equation}
	a(t) = \left(\frac{\Omega_M}{\Omega_\Lambda} \right)^{1/3} \sinh^{2/3} \left(\frac{3}{2} \sqrt{\Omega_\Lambda} \, H_0 t \right).
\end{equation}
The same $H(a)$ is used in every geometry, so that differences between the resulting maps arise only from the spatial geometry in the Ricci rotation coefficients and not from the background expansion. The neglect of radiation is a good approximation for $z \lesssim 1000$ but introduces an error of order $10$--$20\%$ in $H$ at $z \simeq 1200$.

Secondly, we used
\begin{equation}
    J_A = \int (p_{AB}N^B + \hat{p}_{AB} \Bar{N}^B) \frac{d\Omega'}{4\pi}
\end{equation}
where $N^B = N^B (\theta', \phi')$, $\Bar{N}^B = \Bar{N}^B (\theta', \phi')$; and $\text{p}_\text{AB}$ and $\hat{\text{p}}_\text{AB}$ are the scattering 2 $\times$ 2 matrices given by \cite{sung2011temperature}. Note that the Einstein summation convention is applicable only over B and not A(= 0, 2). Also, the emission term $J_A$ contains only harmonics up to $l=2$, since all other terms vanish in virtue of the orthogonality relations for spherical harmonics. This means that the radiation modes with l $\leq$ 2 are damped as well as re-radiated by Thomson scattering, while higher order modes l $>$ 2 are only damped.

%+++++++++++++++++++++++++++++++++++++++++++++++++++++++++++++++++++++++++++++++++++++++++++++++++++++++++++++++++++++++++++++++++++++++++++++++++++++++++++++++++++++++++++++
\subsection{Transfer equations} \label{transfer}
The Boltzmann equation \eqref{boltz} describes the statistical distribution of one particle in a fluid, and is used to study how a fluid transports physical quantities. The angular dependence of polarized radiation fields may be represented by an expansion in spin-zero and spin-two spherical harmonics, so that the expansion is equivalent to a polynomial expansion on the unit sphere \cite{dautcourt1978polarized}. We thus expand the distribution function in \eqref{boltz} in terms of multipole components:
\begin{equation}
\begin{split}
    &N^0 = N^0_0 + N^0_i \, \hat{k}^i + N^0_{ij} \, \hat{k}^{ij} + ....\\
    &N^2 = N^2_{ij}m^{ij} + ...
\end{split}
\end{equation}
where $k^i$ expresses the three-dimensional ray direction given by equation \eqref{ray}, $k^{ij} = k^i k^j - \delta^{ij}/3$, and $m^{ij} = m^i m^j$, where $m^i$ \& $m^j$ are given by equation \eqref{mi}. The number of indices of the k and m polynomials characterize the multipole order of the corresponding contributions to anisotropy and polarization. In terms of spin polynomials, the scattering matrix for Thomson scattering becomes very simple. From \cite{rybicki1979radiative}, the Planck specific intensity $I_\nu (T)$ is given by
\begin{equation} \label{intensity}
	I_\nu (T) = \frac{2 h \nu^3}{c^2} \, \frac{1}{e^{h \nu / k_B T} - 1} \simeq \frac{2 \nu^2 k_B T}{c^2} \quad \text{for} \, \frac{h \nu}{k_B T} << 1
\end{equation}
yielding the antenna (or Rayleigh-Jeans) temp. ($T_A$) as
\begin{equation}
	T_A = \frac{c^2}{2 k_B \nu^2} I_\nu (T)
\end{equation}
which then yields, for a fixed detector frequency,
\begin{equation} \label{RJ}
	\Delta I_\nu (T) = \frac{2 k_B \nu^2}{c^2} \, \Delta T_A
\end{equation}
For standard CMB thermodynamic temperature, using $T = T_\text{CMB}$ in general eq. \eqref{intensity} yields, again for a fixed detector frequency (where $m = h \nu / k_B T_0$, $T_0$ being the isotropic background temperature (2.725 K)),
\begin{equation} \label{CMB}
	\Delta I_\nu (T) = \frac{2 k_B \nu^2}{c^2} \, \frac{m^2 \, e^m}{(e^m - 1)^2} \, \Delta T_\text{CMB}
\end{equation}
From eqs. \eqref{RJ} and \eqref{CMB}, we obtain
\begin{equation}
	\Delta T_\text{CMB} = \frac{(e^m - 1)^2}{m^2 \, e^m} \, \Delta T_A
\end{equation}
For $\nu$ = 150 GHz, we thus get the useful conversion for our final results,
\begin{equation}
	\Delta T_\text{CMB} = 1.736 \, \Delta T_A
\end{equation}
We write our equations for the evolution of the components of the distribution function as
\begin{equation} \label{maint}
\begin{split}
    &\dot{N}^0_0 - \frac{1}{3} \Gamma^0_{kk}  N^0_0 - \frac{2}{15}\zeta \Gamma^0_{kl} N^0_{kl} - \frac{1}{3}\Gamma^k_{ll} N^0_k - \frac{2}{5}\Gamma^0_{kl}N^0_{kl} = -\tau \, i \, \text{Im}\{N^0_0\} \\
    &\dot{N}^0_i + (\hat{A}^k_i \zeta + \hat{B}^k_i)N^0_k + \hat{C}^{kl}_i N^0_{kl} = -\tau \, \left(\text{Re}\{N^0_i\} + \frac{2}{3} \, i  \, \text{Im}\{N^0_i\} \right)\\
    &\dot{N}^0_{ij} + \hat{E}_{ij}\zeta N^0_0 + (\hat{D}^k_{ij}\zeta + \hat{H}^k_{ij})N^0_k + (\hat{F}^{kl}_{ij} \zeta + \hat{G}^{kl}_{ij})N^0_{kl} = -\tau \, \left(\frac{9}{10} \, \text{Re}\{N^0_{ij}\} + \frac{3}{10} \, \text{Re}\{N^2_{ij}\} + i \, \text{Im}\{N^0_{ij}\}\right)\\
    &\dot{N}^2_{ij} - \frac{1}{3}\Gamma^0_{kk}\zeta N^2_{ij} + \hat{K}^{kl}_{ij} N^2_{kl} = -\tau \, \left(\frac{1}{5} \, \text{Re}\{N^0_{ij}\} + \frac{2}{5} \, \text{Re}\{N^2_{ij}\} + i \, \text{Im}\{N^2_{ij}\}\right)
\end{split}
\end{equation}
where
\begin{equation} \label{energy}
    \zeta \equiv \frac{\partial (\text{ln} \, N^0_0)}{\partial (\text{ln} \, \epsilon)}
\end{equation}
%One can use chain rule to isolate the real \& imaginary components of $N^0 _0 = R^0 _0 + i \, I^0_0$ and the photon energy $\epsilon$ as functions of time, where for $d\epsilon/dt$ we used \cite[Eq. (2.26)]{dodelson2003modern} along with equation \eqref{p}. However, we note that the photon monopole component $N^0 _0$ represents the average, isotropic background absolute temperature of about 2.7 K, thus making it an essentially real quantity and we can safely set $\Im \{N^0 _0\} = 0$. Considering then the radiation field be described by a Planck distribution at all times, 
%
%From eqs. \eqref{NV} \& \eqref{intensity}, we obtain
%\begin{equation} \label{Planck}
%	N^0 _0 = \frac{c^2}{2 h \nu^3} \, I_\nu (T_0) = \frac{1}{e^x - 1}
%\end{equation}
%Substituting equation \eqref{Planck} into equation \eqref{energy}, we obtain $\zeta$ purely in terms of the ratio $\left(\epsilon / T \right)$, as
%\begin{equation}
%	\zeta = -\frac{x \, e^x}{e^x - 1} 
%\end{equation}
%The ratio $x = \epsilon / k_B T$ remains conserved in an expanding universe. For $\nu_0 = 150 \, \text{GHz}$ and $T_0 = 2.725 \, \text{K}$, x = 2.643 (cf. the Wien peak at $x \simeq 2.82$). Thus, $\zeta = -2.845$, a real and negative quantity, as is required for a thermal spectrum.

The coefficients arising in the transfer equation are given by
\begin{align*}
    \hat{A}^k_i &= \frac{1}{5} (\Gamma^0_{ik} - \Gamma^0_{ki} + \Gamma^0_{ll} \delta_{ik}) \\
    \hat{B}^k_i &= -\Gamma^k_{0i} + \frac{1}{5} (\Gamma^0_{ki} - 4\Gamma^0_{ik} + \Gamma^l_{l0} \delta_{ik}) \\
    \hat{C}^{kl}_i &= -\frac{2}{5} (\Gamma^k_{li} + \Gamma^k_{mm} \delta_{il} + 3\Gamma^k_{00} \delta_{li}) \\
    \hat{D}^k_{ij} &= \frac{1}{3} \Gamma^0_{0l} \delta_{ij} - \frac{1}{2} \Gamma^0_{0i} \delta_{lj} - \frac{1}{2} \Gamma^0_{0j} \delta_{li} \\
    \hat{E}_{ij} &= \frac{1}{3} \Gamma^0_{ll} \delta_{ij} - \frac{1}{2} \Gamma^0_{ij} - \frac{1}{2} \Gamma^0_{ji} \\
    \hat{F}^{kl}_{ij} &= \frac{2}{21} \Gamma^0_{kl} \delta_{ij} - \frac{1}{7} (\delta_{ki} \delta_{lj} \Gamma^0_{mm} + \Gamma^0_{ki} \delta_{jl} + \Gamma^0_{kj} \delta_{il}) \\
    \hat{G}^{kl}_{ij} &= \frac{2}{21} \Gamma^0_{kl} \delta_{ij} + \frac{2}{7} (\Gamma^0_{ki} \delta_{jl} + \Gamma^0_{kj} \delta_{il}) - \Gamma^k_{0i} \delta_{lj} - \Gamma^k_{0j} \delta_{li} -\frac{5}{7} (\Gamma^0_{ik} \delta_{jl} + \Gamma^0_{jk} \delta_{il}) + \frac{2}{7} \Gamma^0_{mm} \delta_{ik} \delta_{jl} \\
    \hat{H}^k_{ij} &= -\frac{1}{2} \Gamma^k_{ij} - \frac{1}{2} \Gamma^k_{ji} + \frac{1}{3} \Gamma^k_{mm} \delta_{ij} + \frac{1}{2} \Gamma^j_{00} \delta_{ik} + \frac{1}{2} \Gamma^i_{00} \delta_{jk} - \frac{1}{3} \Gamma^k_{00} \delta_{ij} \\
    \hat{K}^{kl}_{ij} &= -\frac{2}{9} \Gamma^0_{kl} \delta_{ij} + \frac{1}{3} (\Gamma^k_{0i} + \Gamma^k_{i0}) \delta_{lj} + \frac{1}{3} (\Gamma^j_{0k} + \Gamma^j_{k0}) \delta_{li} + \frac{i}{3} \delta_{ki} \delta_{lj} \Gamma^s_{tr} \epsilon_{rst}
\end{align*}
We codify all our equations purely in terms of $\kappa$ and not in L. The patterns that are produced, therefore, depend both on the background cosmology and the initial conditions. Figure \ref{polarization_mechanism} summarizes the general mechanism of the generation of polarization amplitudes from gravitational field codified in the set of equations \eqref{maint}. It is to be noted that the presented form of Thurston geometries incorporate a zero shear tensor due to the presence of a single scale factor a(t), but a non-zero intrinsic geometrical shear due to curvature.
\begin{figure}[htbp]
	\centering
	\begin{tikzpicture}[
		node distance=1.8cm,
		block_orange/.style={
			rectangle, 
			draw=orange!90!black, 
			thick, 
			text width=4.5cm, 
			align=center, 
			minimum height=1.2cm,
			text=orange!90!black,
			font=\sffamily
		},
		block_green/.style={
			rectangle, 
			draw=green!60!black, 
			thick, 
			text width=4.5cm, 
			align=center, 
			minimum height=1.2cm,
			text=green!60!black,
			font=\sffamily
		},
		arrow_orange/.style={
			-Stealth, 
			thick, 
			orange!90!black
		},
		arrow_green/.style={
			-Stealth, 
			thick, 
			green!60!black
		}
		]
		
		% Nodes
		\node (grav) [block_orange] {Gravitational Field / Geometry};
		
		\node (quad) [block_green, below=of grav] {Radiation Quadrupole Anisotropy};
		
		\node (pol) [block_green, below=of quad] {Linear Polarization};
		
		% Arrows
		\draw [arrow_orange] (grav) -- node[right, align=left, font=\sffamily\small] {Shear\\$\tilde{E}_{ij} = -\sigma_{ij}$} (quad);
		
		\draw [arrow_green] (quad) -- node[right, font=\sffamily\small] {Thomson scattering} (pol);
		
	\end{tikzpicture}
	\caption{Mechanism of polarization generation from gravitational field}
	\label{polarization_mechanism}
\end{figure}

%+++++++++++++++++++++++++++++++++++++++++++++++++++++++++++++++++++++++++++++++++++++++++++++++++++++++++++++++++++++++++++++++++++++++++++++++++++++++++++++++++++++++++++++
\subsection{Geodesic equations} \label{geodesic}
The equations in \S\ref{transfer} are solved alongside the geodesic equations described in this section for each of our geometries. We start with equation \eqref{p} to write
\begin{equation} \label{decompose}
\begin{split}
    dp^i &= k^i \, d\epsilon + \epsilon \, dk^i \\
    &= k^i \, d\epsilon + \epsilon \left(\frac{\partial k^i}{\partial \theta} d\theta + \frac{\partial k^i}{\partial \phi} d\phi \right) \\
    &= k^i \, d\epsilon + \epsilon \left(a^i d\theta + b^i \, \sin \theta \, d\phi\right)
\end{split}
\end{equation}
where we used equation \eqref{ray}
\begin{equation}
	k^i =
\begin{bmatrix}
	\sin \theta \cos \phi & \sin \theta \sin \phi & \cos \theta
\end{bmatrix}
\end{equation}
along with (see \cite{dautcourt1978polarized})
\begin{subequations}
\begin{gather}
    a^i = \frac{\partial k^i}{\partial \theta} =
    \begin{bmatrix}
    	\cos \theta \cos \phi & \cos \theta \sin \phi & -\sin \theta
    \end{bmatrix}
    \\
    b^i = \frac{1}{\sin \theta} \frac{\partial k^i}{\partial \phi} =
    \begin{bmatrix}
    	-\sin \phi & \cos \phi & 0
    \end{bmatrix}
\end{gather}
\end{subequations}
Using $k_i = \left(k^i \right)^T$, $a_i = \left(a^i \right)^T$ \& $b_i = \left(b^i \right)^T$, since $k^i$, $a^i$ \& $b^i$ are being written in the orthonormal spatial tetrad instead of the coordinate basis, and thus utilizing the orthonormality relations
\begin{subequations}
\begin{gather}
	k^i k_i = a^i a_i = b^i b_i = 1 \\
	k^i a_i = k^i b_i = a^i k_i = a^i b_i = b^i k_i = b^i a_i = 0
\end{gather}
\end{subequations} 
in equation \eqref{decompose}, we obtain
\begin{subequations}
\begin{gather}
	\frac{d\theta}{dt} = \frac{a_i}{\epsilon} \, \frac{dp^i}{dt} \label{geo_1} \\
	\frac{d\phi}{dt} = \frac{b_i}{\epsilon \, \sin \theta} \, \frac{dp^i}{dt} \label{geo_2} 
\end{gather}
\end{subequations}
Using the photon geodesic equation
\begin{equation}
	\frac{dp^i}{d\lambda} = -\Gamma^i _{\beta\gamma} \, p^\beta \, p^\gamma = -\epsilon^2 \, \Bar{\gamma}^i
\end{equation}
where we used the definition of $\Bar{\gamma}$ from equation \eqref{gamma_def}, and noting that
\begin{equation}
	\frac{dp^i}{d\lambda} = \frac{dp^i}{dt} \, \frac{dt}{d\lambda} = \epsilon \frac{dp^i}{dt}
\end{equation}
we have
\begin{equation}
	\frac{dp^i}{dt} = -\epsilon \, \Bar{\gamma}^i
\end{equation}
Then equations \eqref{geo_1}-\eqref{geo_2} take the form
\begin{subequations}
\begin{gather}
	\frac{d\theta}{dt} = -a_i \Bar{\gamma}^i \label{geo_final_1} \\
	\frac{d\phi}{dt} = -\frac{b_i}{\sin \theta} \Bar{\gamma}^i \label{geo_final_2}
\end{gather}
\end{subequations}
along with \cite{sung2011temperature}
\begin{equation} \label{polangle}
	\frac{d\psi}{dt} = -b_l \, \bar{\gamma}^l \, \cot \theta + i \, m^i \, \Bar{m}^k \, \Gamma^k_{li} \, k^l
\end{equation}
where equation \eqref{polangle} represents the change of polarization angle, and $\Bar{m}^k$ represents the complex conjugate of $m^k$. Following equation \eqref{NV}, the Stokes parameter `V' is constructed from the numerical solution $\Im{\left(N^0 \right)}$ of the set of transfer equations \eqref{maint} itself, and not from raw multipoles $N^0$. Thus, `V' already carries the $\left(1/ch^4 \nu^3 \right)$ prefactor (with the physical specific intensities) by definition. Therefore, no additional factor is applied.

Equations \eqref{geo_final_1}, \eqref{geo_final_2} and \eqref{polangle} are solved alongside the set of transfer equations \eqref{maint} to obtain the correct photon kinematics for each of our considered Thurston geometry.

%+++++++++++++++++++++++++++++++++++++++++++++++++++++++++++++++++++++++++++++++++++++++++++++++++++++++++++++++++++++++++++++++++++++++++++++++++++++++++++++++++++++++++++++
\section{Results} \label{four}
In this section, we will present representative examples of the temperature and polarization patterns produced in anisotropic Thurston geometries \eqref{rh2s2} - \eqref{solv}, computed by integrating the system of equations in \S\ref{transfer} and \ref{geodesic}. 

Specifically, our work simulates the redshift evolution of CMB temperature and polarization fluctuations in each anisotropic Thurston background. The expansion history is obtained from the Friedmann equation for pressureless matter, a cosmological constant and spatial curvature,
\begin{equation}
	H(a)/H_0 = \sqrt{\Omega_m a^{-3} + \Omega_k a^{-2} + \Omega_\Lambda},
\end{equation}
with $\Omega_m = 0.315$, $\Omega_k = 0.044$ \cite{awwad2024large} and $\Omega_\Lambda = 1 - \Omega_m - \Omega_k = 0.641$. Radiation is neglected, since the simulated epoch ($z = 1200$ to $0$) is dominated by matter and dark energy. In the flat limit ($\Omega_k = 0$) this reduces to the closed form $a(t) \propto \sinh^{2/3}\!\left(\tfrac{3}{2}\sqrt{\Omega_\Lambda}\,H_0 t\right)$. The same background is used in every geometry, so that differences between the maps arise only from the spatial geometry. The Thomson scattering rate $\tau(z) = n_e\sigma_T c$ \cite{sung2011temperature, dodelson2003modern} is not modeled analytically. It is computed from the opacity of the Boltzmann code \texttt{CAMB}, which includes the full hydrogen and helium recombination history and reionization with $\tau_{\rm reio} = 0.054$. It is tabulated on a grid uniformly spaced in $\ln(1+z)$ for $0 \le z \le 3000$ and interpolated during the integration.

Our core methodology is listed in the following key steps. All theoretical expressions are in natural units.
\begin{enumerate}[label=\Roman*.]
	\item For each line of sight, the photon distribution is evolved by integrating a stiff system of coupled ordinary differential equations (ODEs). It tracks the real and imaginary parts of the multipoles ($R^0_\mu, I^0_\mu, R^0_{ij}, I^0_{ij}, R^2_{ij}, I^2_{ij}$), which gives 44 moment equations, together with the geodesic variables $(\psi, \theta, \phi, \chi$ and, where needed, the Cartesian coordinates$)$. The total number of equations is 49 to 52, depending on the geometry. The radiation field is taken to be Planckian, with a constant spectral index $\zeta$. Initial quadrupole perturbations with $m = 0, 1, 2$ are seeded as in \S\ref{form}, and their contributions are summed. At every function evaluation, symmetry $\left(N^{0,2}_{ij} = N^{0,2}_{ji}\right)$ and tracelessness $\left(\delta^{ij}N^{0,2}_{ij} = 0\right)$ are enforced on a copy of the state vector, which prevents numerical drift of the off-diagonal components without corrupting the solver's Jacobian estimate.
	
	\item The sky is discretized into 12,288 pixels using \texttt{HEALPix} ($N_{\rm side} = 32$). The scale factor $a$ is used as the independent variable, with $d/dt = aH(a)\,d/da$, and the comoving distance is $\chi(a) = \int_a^1 da'/[a'^2H(a')]$. Using Python's \texttt{multiprocessing}, each pixel is integrated independently from $z = 1200$ to $0$ with the implicit \texttt{Radau} method, which is suited to the stiff tightly coupled regime ($\tau/H \gg 1$). The tolerances are $\mathrm{rtol} = 10^{-6}$, with component-wise absolute tolerances: $10^{-13}$ for the multipoles, $10^{-10}$ for the angles and $10^{6}$ for the coordinates.
	
	\item For each line of sight, the multipole tensors are projected onto the local orthonormal basis $(\hat{e}_\theta, \hat{e}_\phi)$ defined at each sky position $(\theta,\phi)$ as
	\begin{equation}
		\hat{e}_\theta = (\cos\theta\cos\phi,\ \cos\theta\sin\phi,\ -\sin\theta), \quad \hat{e}_\phi = (-\sin\phi,\ \cos\phi,\ 0).
	\end{equation}
	The temperature fluctuation is obtained by contracting the quadrupole multipoles with $k^ik^j - \delta^{ij}/3$. In this frame, $Q > 0$ denotes polarization preferentially along the meridional ($\hat{e}_\theta$) direction and $Q < 0$ along the zonal ($\hat{e}_\phi$) direction, after a rotation by $2\psi$, where $\psi$ is the polarization-plane angle accumulated along the null geodesic.
	
	\item The work produces a grid of Mollweide projections of $\Delta T$, $P$, $Q$ and $U$ at $z = 1200$, $1100$, $550$ and $0$. For visual clarity across epochs, the colourbar limits are recomputed at each redshift from the \emph{maximum} absolute value of the map.
\end{enumerate}
It has become conventional to decompose the polarized component of the CMB into parity-classified modes, where even modes are termed $E$-modes and odd modes $B$-modes. The latter are of particular interest in standard inflationary cosmology because they cannot be sourced by scalar perturbations at linear primordial order in standard cosmology - although secondary and nonlinear effects, such as gravitational lensing, can generate them \cite{kamionkowski1997statistics, hu1997cmb}. As discussed in several recent studies, anisotropic and non-standard geometries can also source distinct $B$-mode signatures \cite{sung2009polarized, pontzen2007bianchi, pontzen2009rogues, pontzen2011linearization}. While an explicit $E/B$ (along with $EB$ and $TB$) decomposition is a powerful tool for directly investigating parity questions, a complete gallery of the $E$- and $B$-mode behaviors for these specific models has already been presented by Pontzen \cite{pontzen2009rogues}. To complement rather than duplicate those findings, we present our primary results in terms of the real-space $Q$ and $U$ Stokes parameters, which contain equivalent linear-polarization information and directly illustrate the spatial evolution of the tensor patterns.

\begin{figure}[p]
	\centering
	\includegraphics[width=\textwidth,
	height=\dimexpr\textheight-5\baselineskip\relax,
	keepaspectratio]{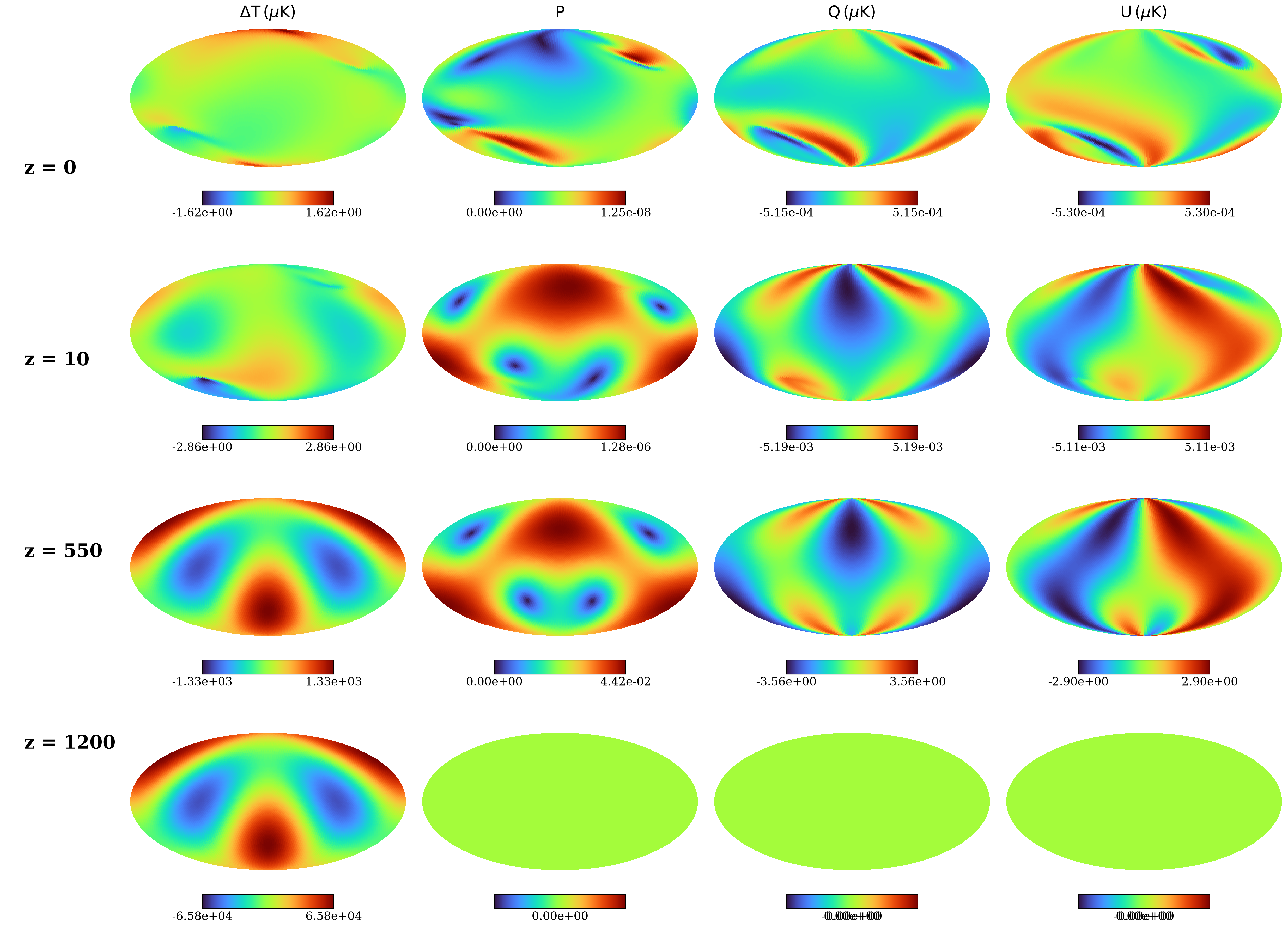}
	\caption{Temperature and polarization patterns for $\mathbb{R}^3$ geometry}
	\label{R3}
\end{figure}

\begin{figure}[p]
	\centering
	\includegraphics[width=\textwidth,
	height=\dimexpr\textheight-5\baselineskip\relax,
	keepaspectratio]{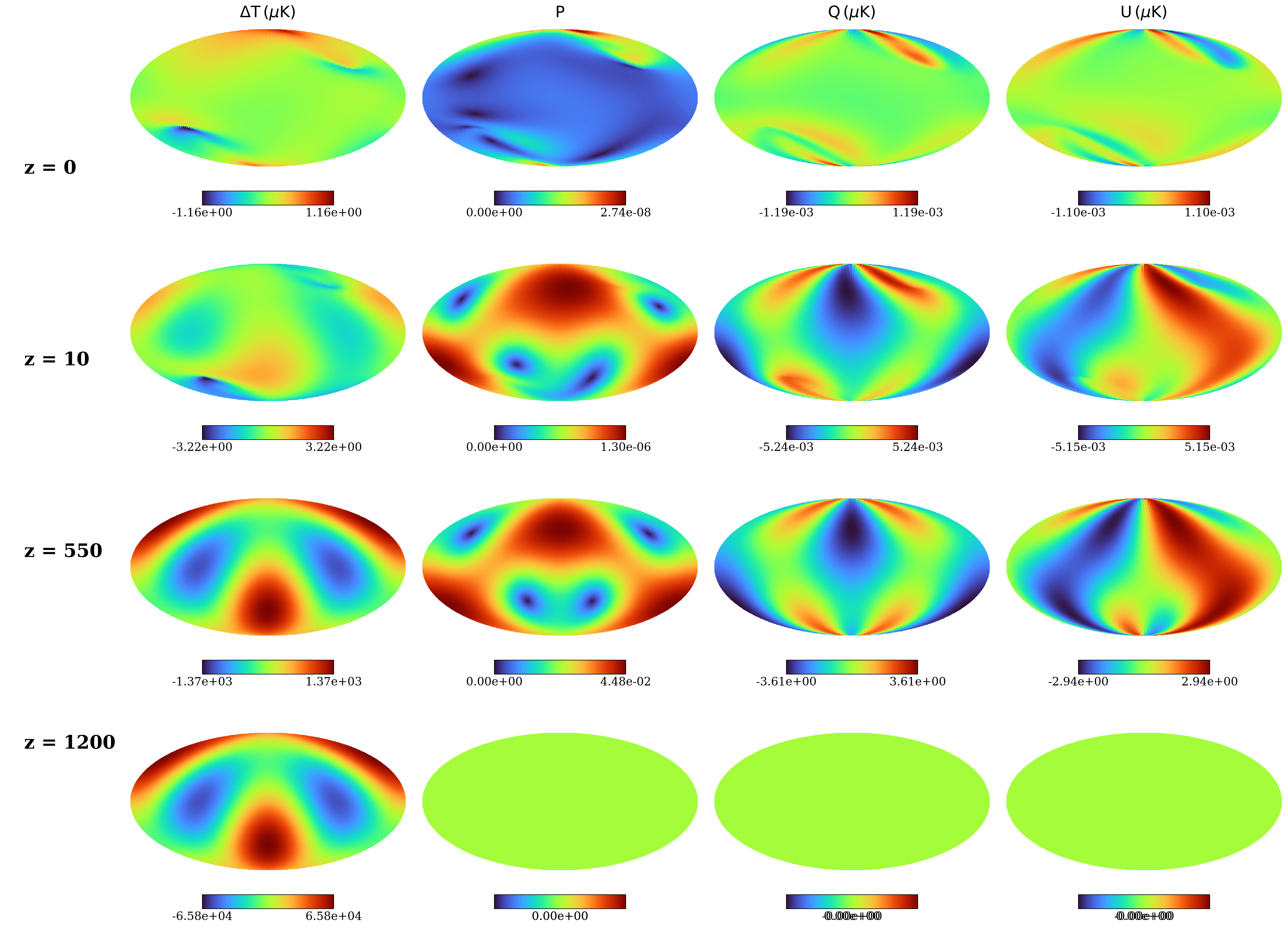}
	\caption{Temperature and polarization patterns for $S^3$ geometry}
	\label{S3}
\end{figure}

\begin{figure}[p]
	\centering
	\includegraphics[width=\textwidth,
	height=\dimexpr\textheight-5\baselineskip\relax,
	keepaspectratio]{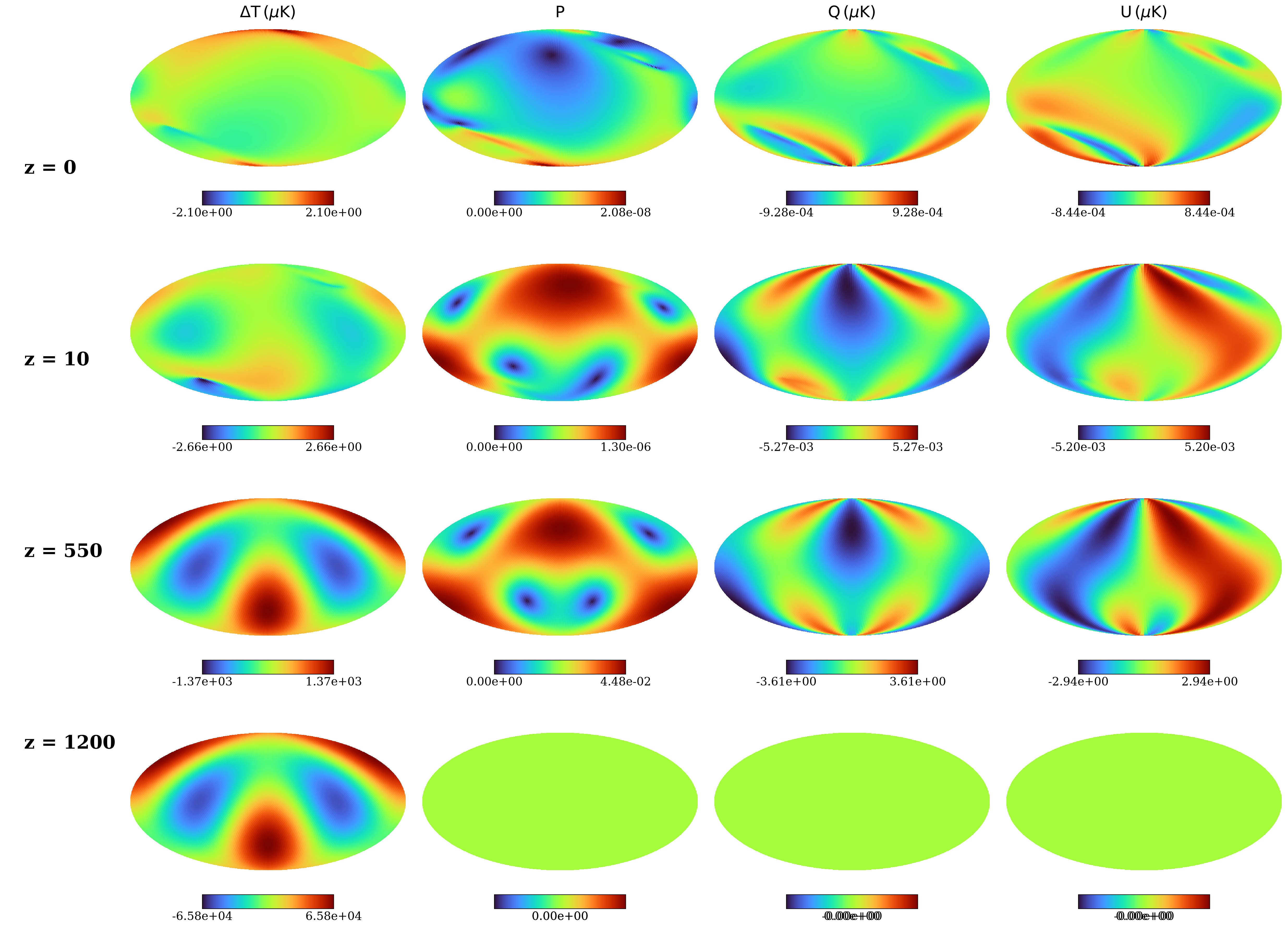}
	\caption{Temperature and polarization patterns for $\mathbb{H}^3$ geometry}
	\label{H3}
\end{figure}

\begin{figure}[p]
	\centering
	\includegraphics[width=\textwidth,
	height=\dimexpr\textheight-5\baselineskip\relax,
	keepaspectratio]{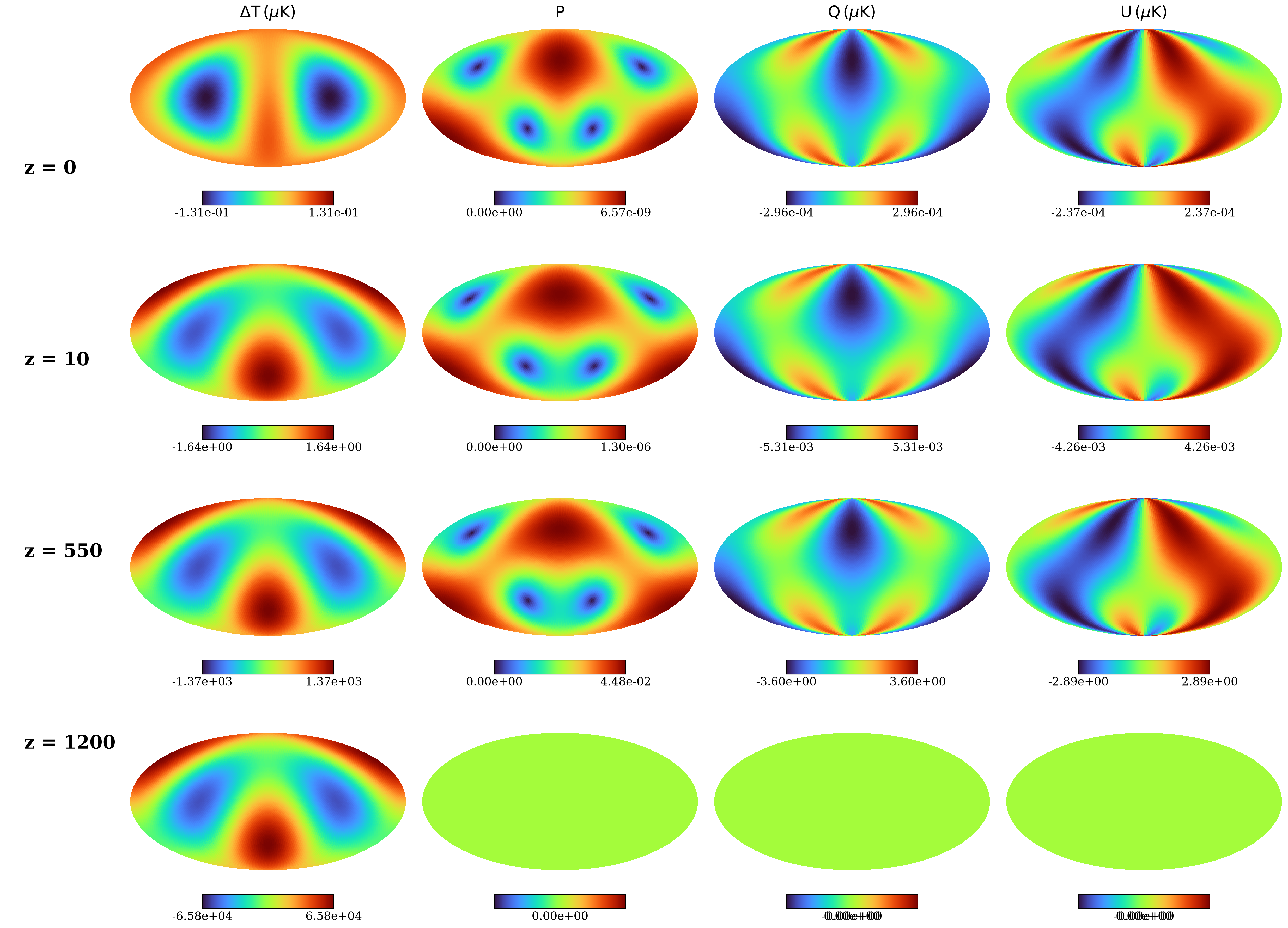}
	\caption{Temperature and polarization patterns for $\mathbb{R} \times S^2$ geometry}
	\label{RS2}
\end{figure}

\begin{figure}[p]
	\centering
	\includegraphics[width=\textwidth,
	height=\dimexpr\textheight-5\baselineskip\relax,
	keepaspectratio]{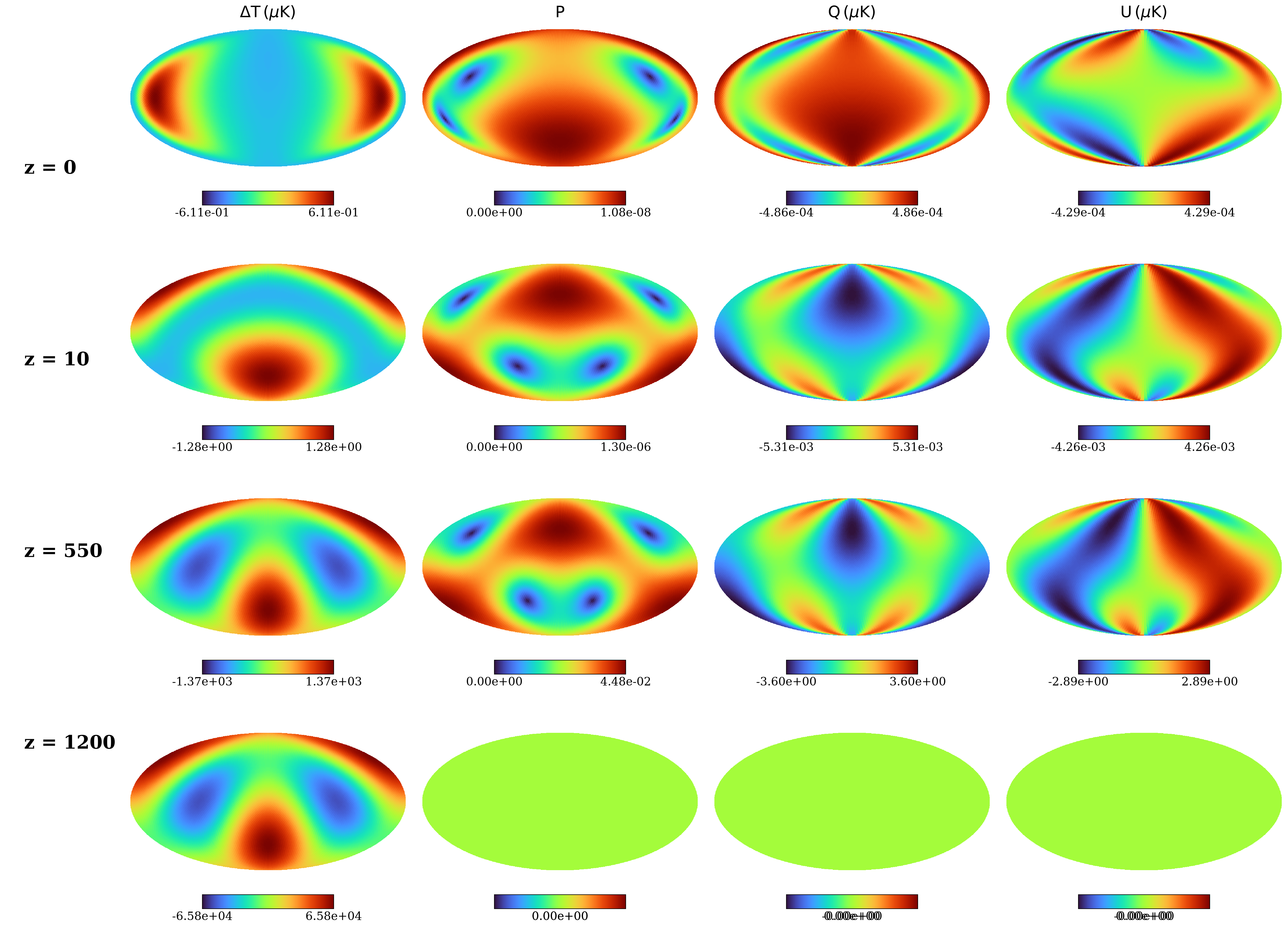}
	\caption{Temperature and polarization patterns for $\mathbb{R} \times \mathbb{H}^2$ geometry}
	\label{RH2}
\end{figure}

\begin{figure}[p]
	\centering
	\includegraphics[width=\textwidth,
	height=\dimexpr\textheight-5\baselineskip\relax,
	keepaspectratio]{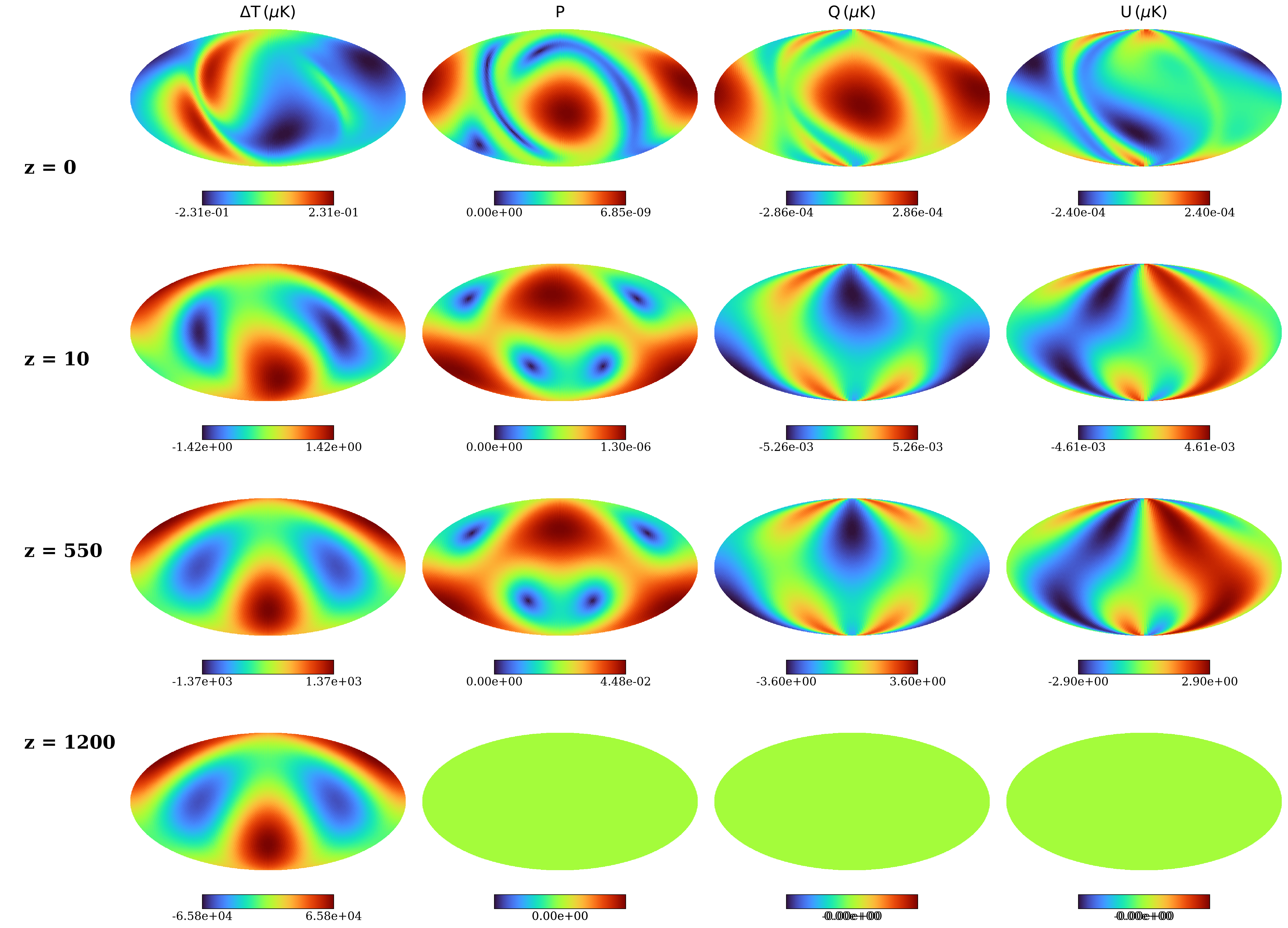}
	\caption{Temperature and polarization patterns for $\widetilde{U \left(\mathbb{H}^2 \right)}$ geometry}
	\label{UH2}
\end{figure}

\begin{figure}[p]
	\centering
	\includegraphics[width=\textwidth,
	height=\dimexpr\textheight-5\baselineskip\relax,
	keepaspectratio]{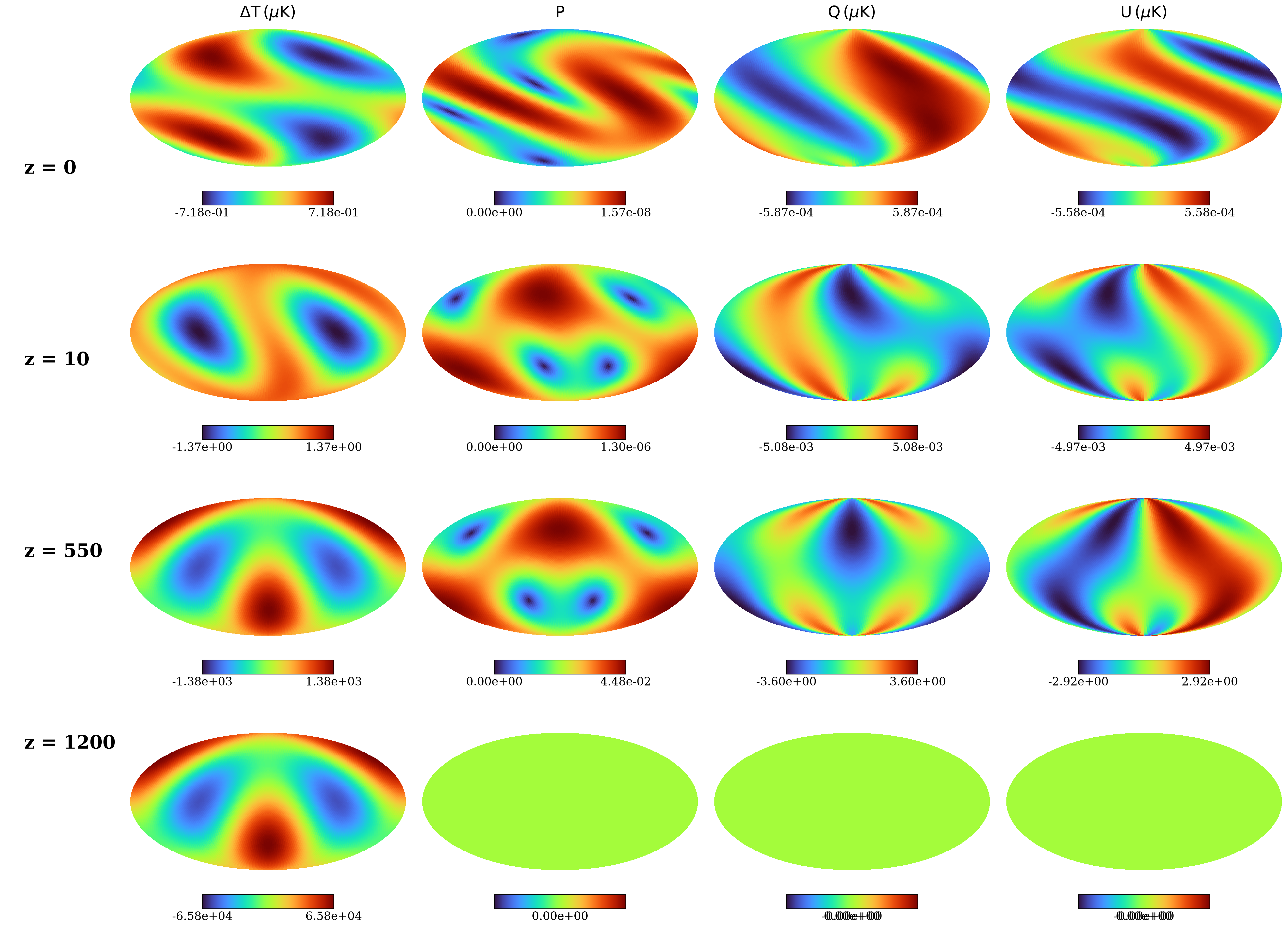}
	\caption{Temperature and polarization patterns for Nil geometry}
	\label{Nil}
\end{figure}

\begin{figure}[p]
	\centering
	\includegraphics[width=\textwidth,
	height=\dimexpr\textheight-5\baselineskip\relax,
	keepaspectratio]{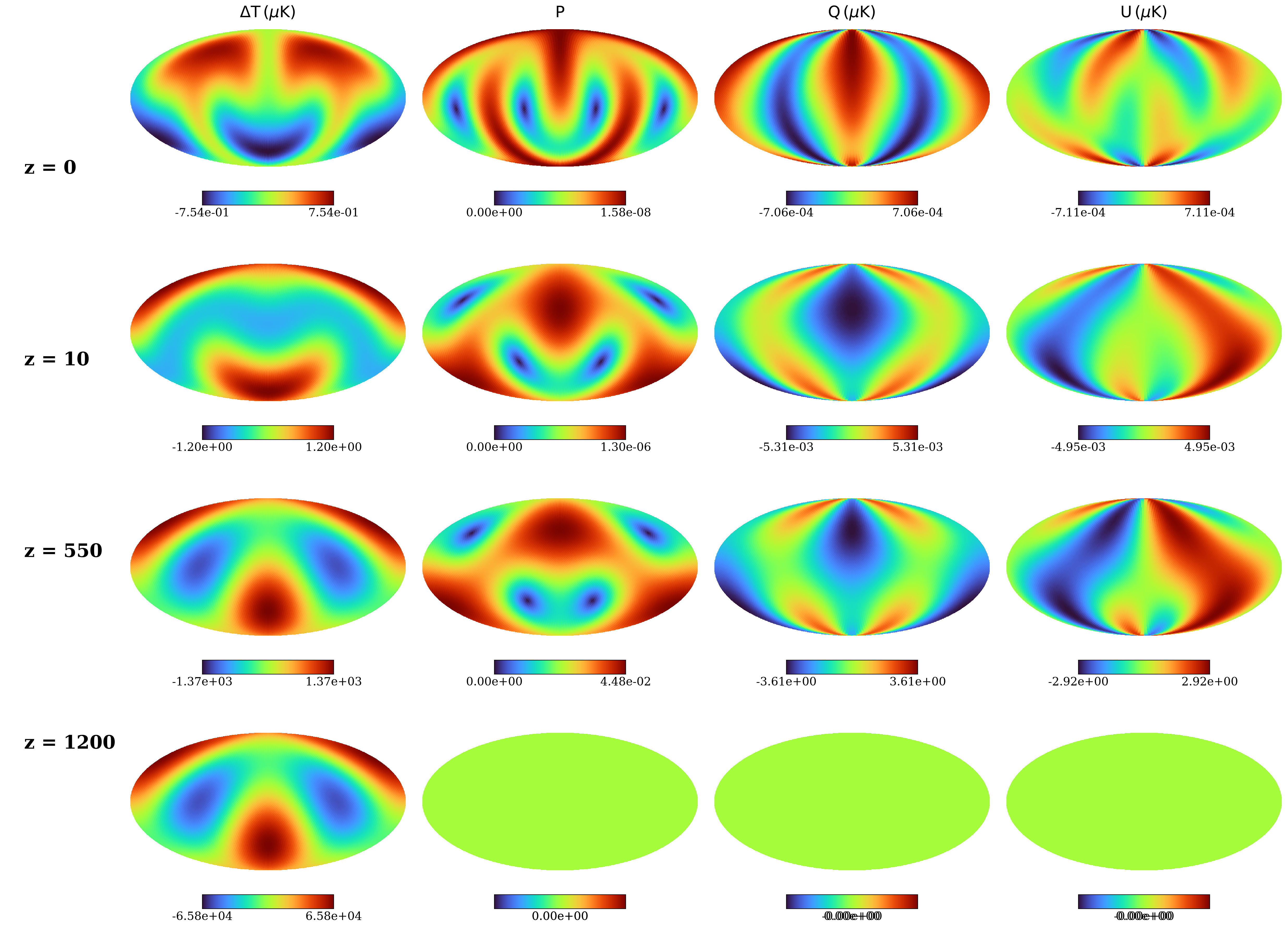}
	\caption{Temperature and polarization patterns for Solv geometry}
	\label{Solv}
\end{figure}

\begin{figure}[p]
	\centering
	\captionsetup[subfigure]{font=small, skip=2pt}
	
	\begin{subfigure}[t]{0.48\textwidth}
		\centering
		\includegraphics[width=\linewidth]{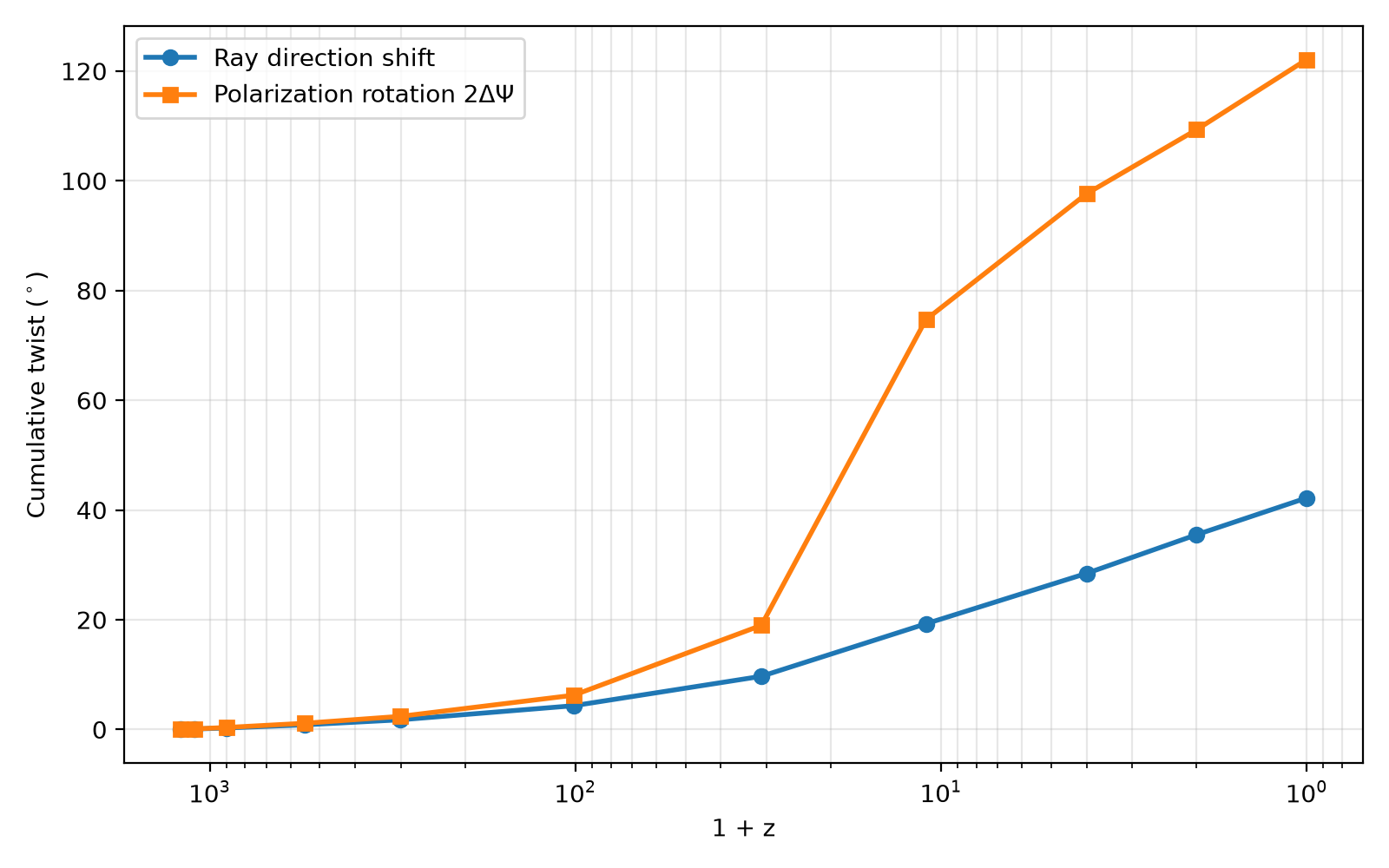}
		\caption{$\mathbb{R}^3$}
		\label{R3_twist}
	\end{subfigure}\hfill
	\begin{subfigure}[t]{0.48\textwidth}
		\centering
		\includegraphics[width=\linewidth]{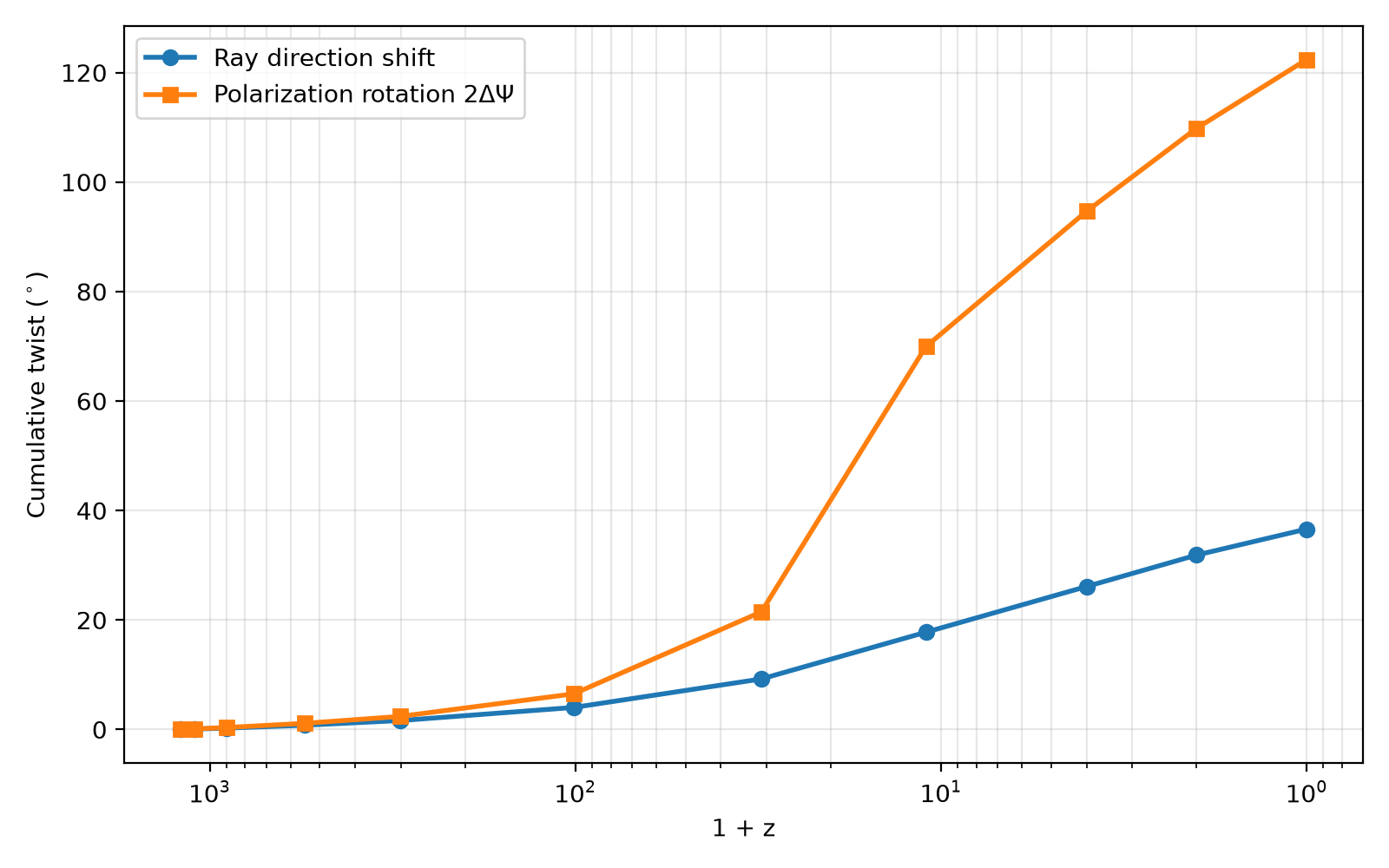}
		\caption{$S^3$}
		\label{S3_twist}
	\end{subfigure}
	
	\vspace{4pt}
	
	\begin{subfigure}[t]{0.48\textwidth}
		\centering
		\includegraphics[width=\linewidth]{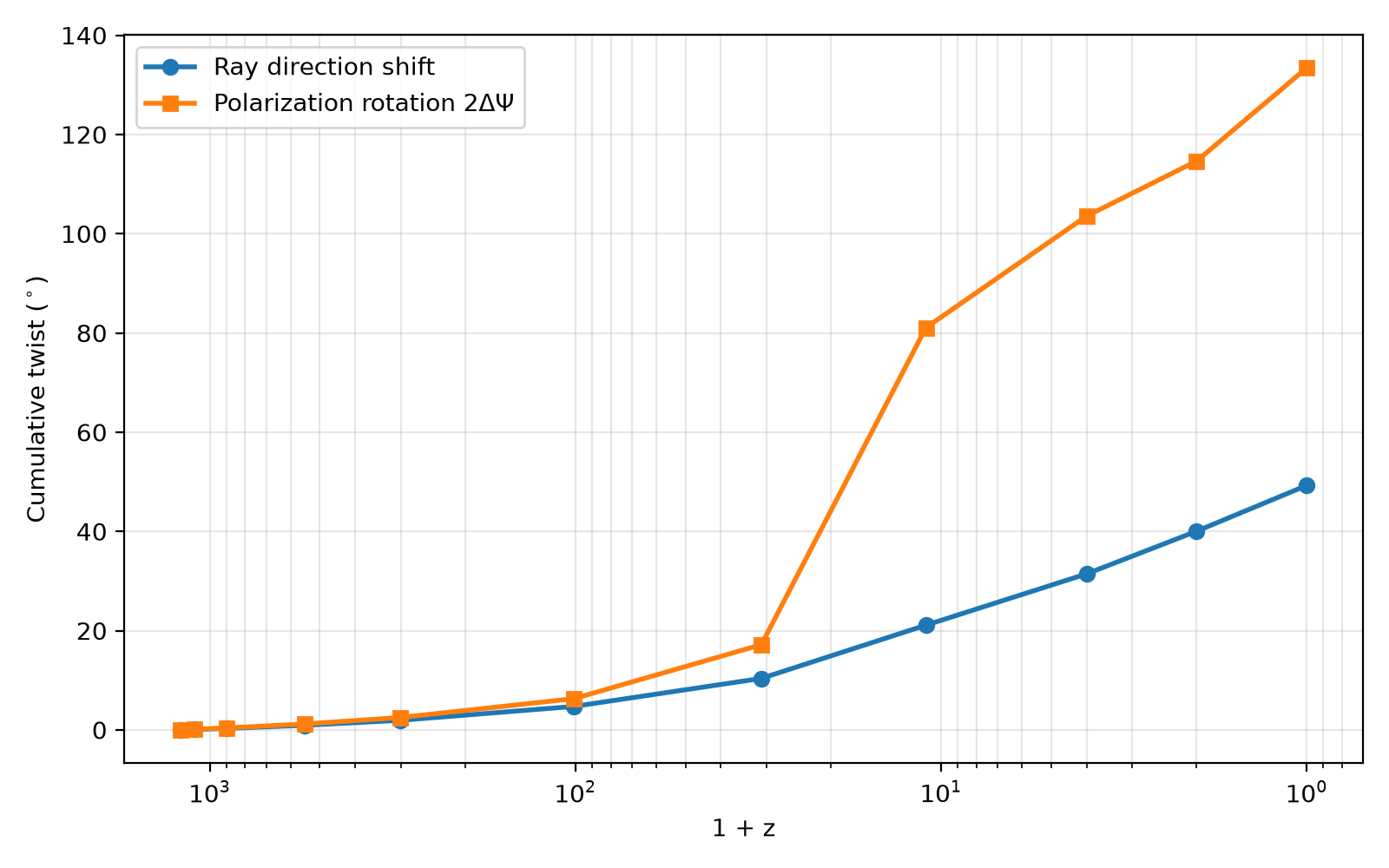}
		\caption{$\mathbb{H}^3$}
		\label{H3_twist}
	\end{subfigure}\hfill
	\begin{subfigure}[t]{0.48\textwidth}
		\centering
		\includegraphics[width=\linewidth]{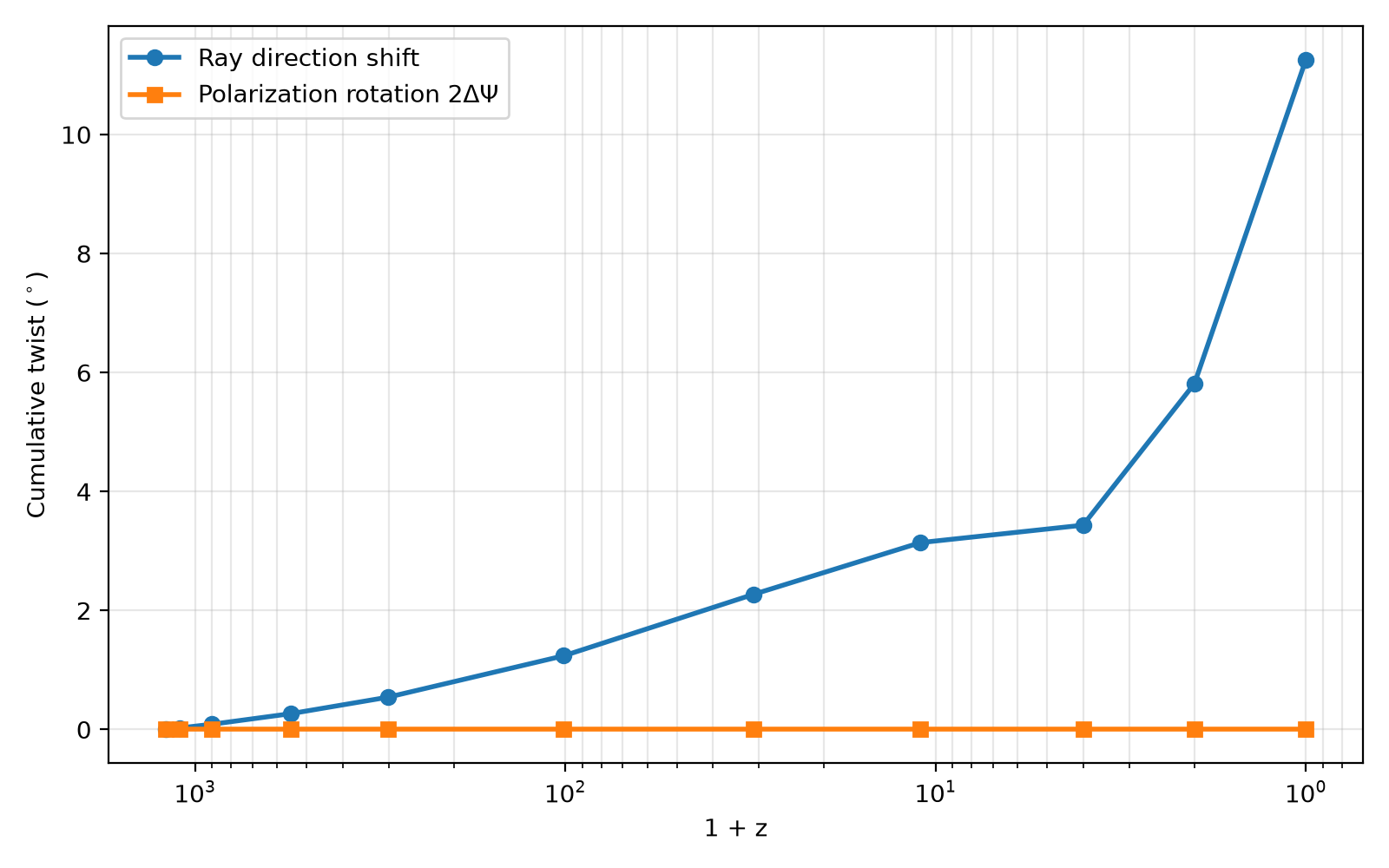}
		\caption{$\mathbb{R} \times S^2$}
		\label{RS2_twist}
	\end{subfigure}
	
	\vspace{4pt}
	
	\begin{subfigure}[t]{0.48\textwidth}
		\centering
		\includegraphics[width=\linewidth]{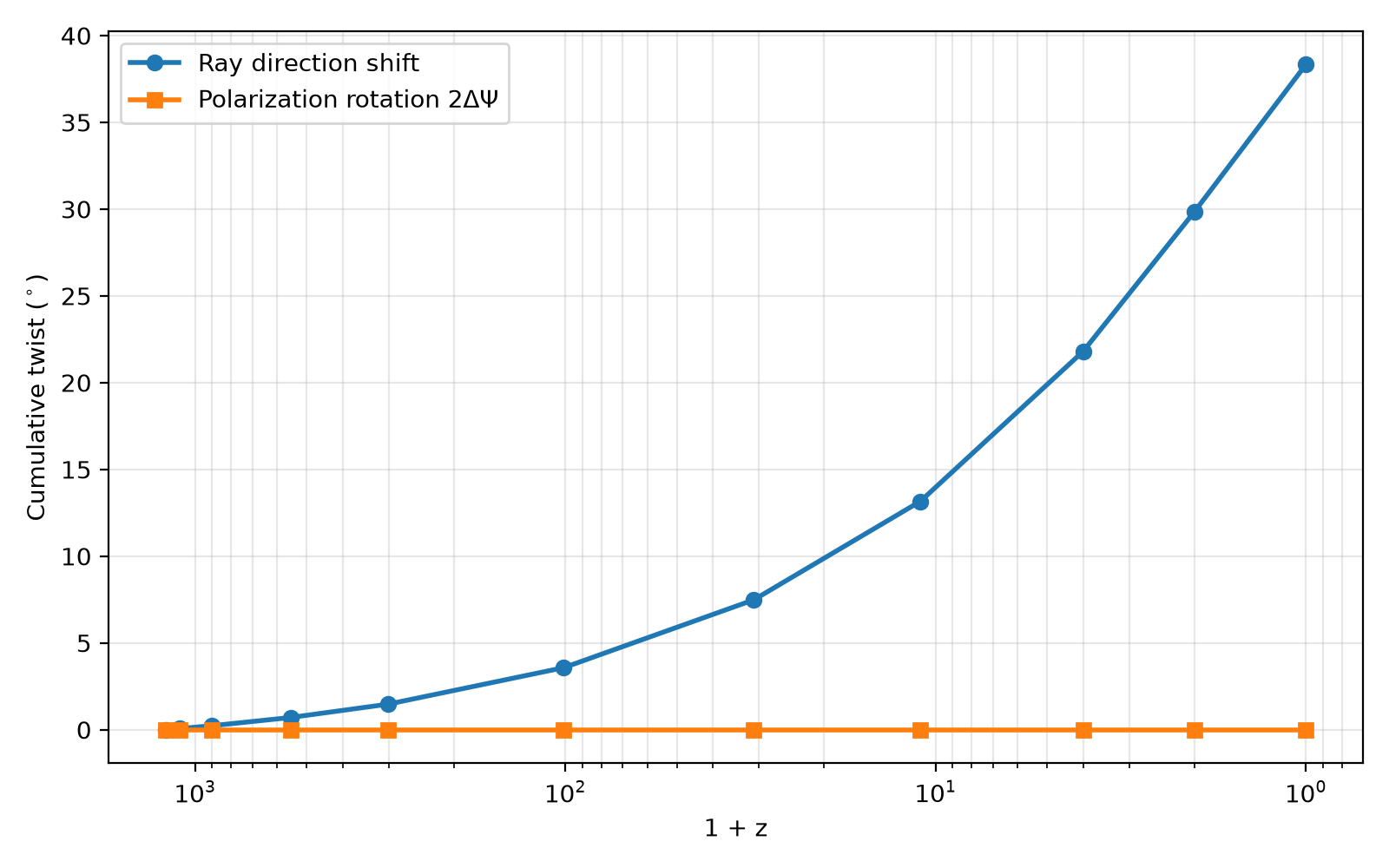}
		\caption{$\mathbb{R} \times \mathbb{H}^2$}
		\label{RH2_twist}
	\end{subfigure}\hfill
	\begin{subfigure}[t]{0.48\textwidth}
		\centering
		\includegraphics[width=\linewidth]{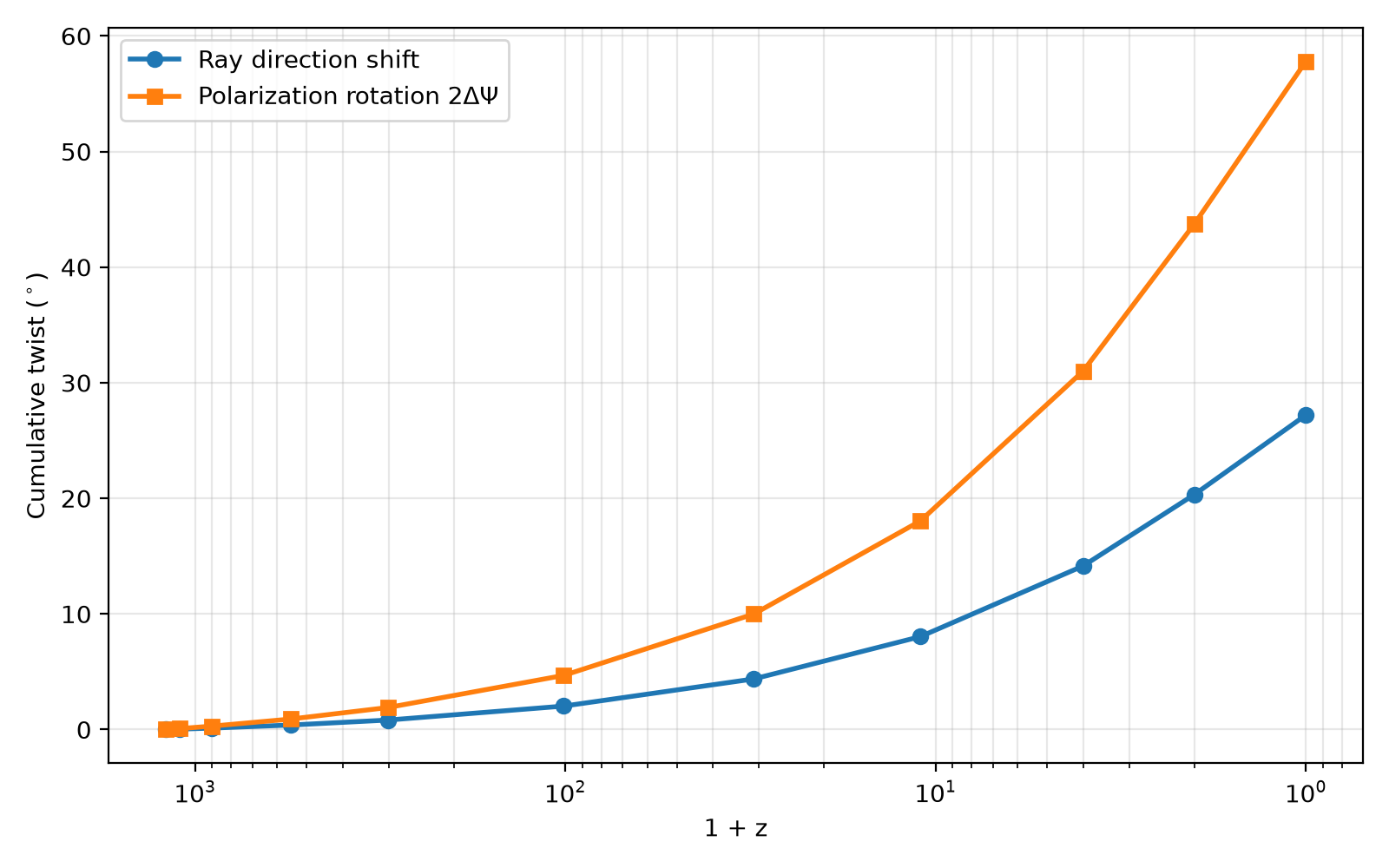}
		\caption{$\widetilde{U \left(\mathbb{H}^2 \right)}$}
		\label{UH2_twist}
	\end{subfigure}
	
	\vspace{4pt}
	
	\begin{subfigure}[t]{0.48\textwidth}
		\centering
		\includegraphics[width=\linewidth]{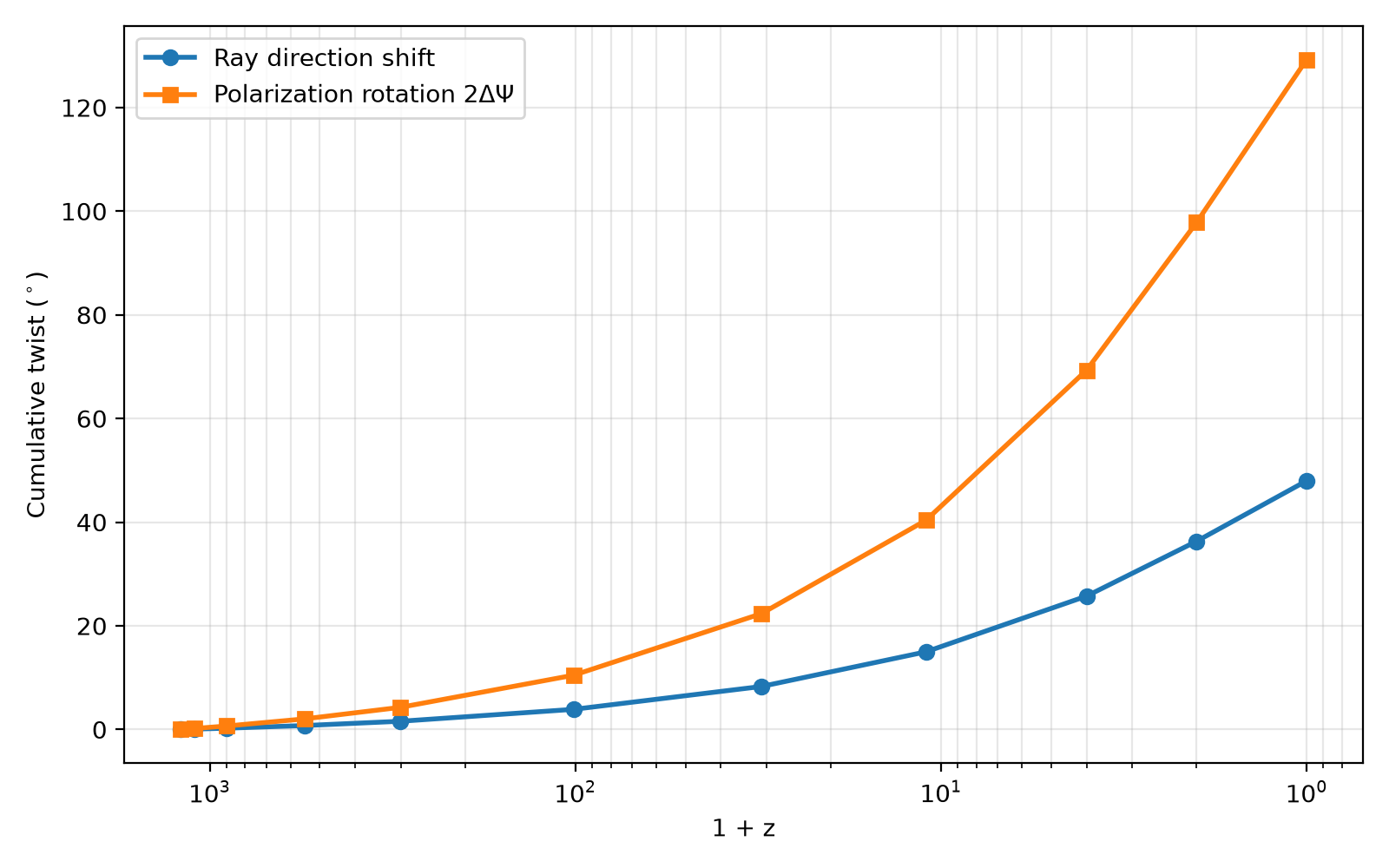}
		\caption{Nil}
		\label{Nil_twist}
	\end{subfigure}\hfill
	\begin{subfigure}[t]{0.48\textwidth}
		\centering
		\includegraphics[width=\linewidth]{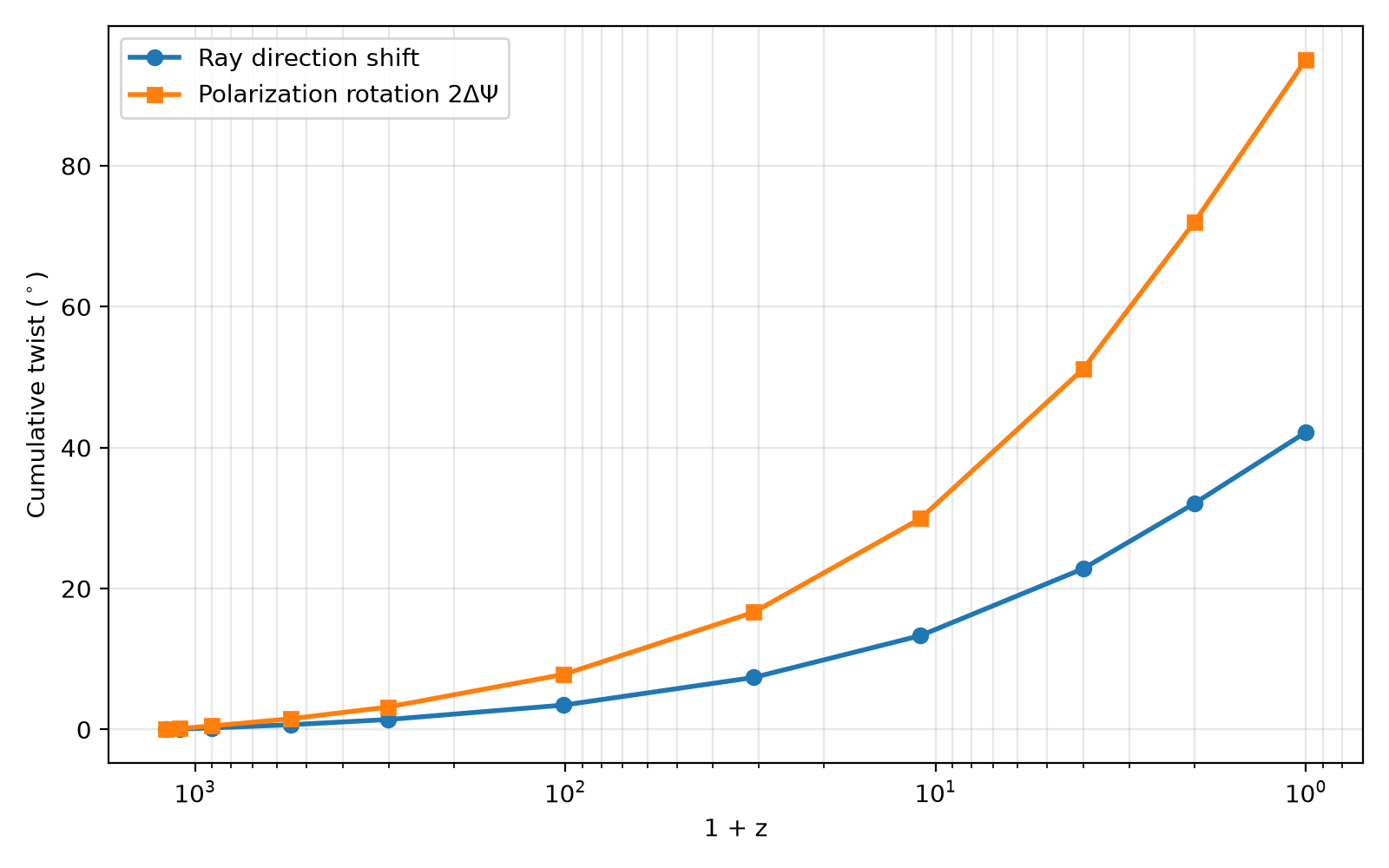}
		\caption{Solv}
		\label{Solv_twist}
	\end{subfigure}
	
	\caption{Twist buildup against redshift in Thurston geometries, explaining the rapid change in signals at low redshifts.}
	\label{twist}
\end{figure}

\begin{figure}[!t]
	\centering
	\includegraphics[width=0.7\linewidth]{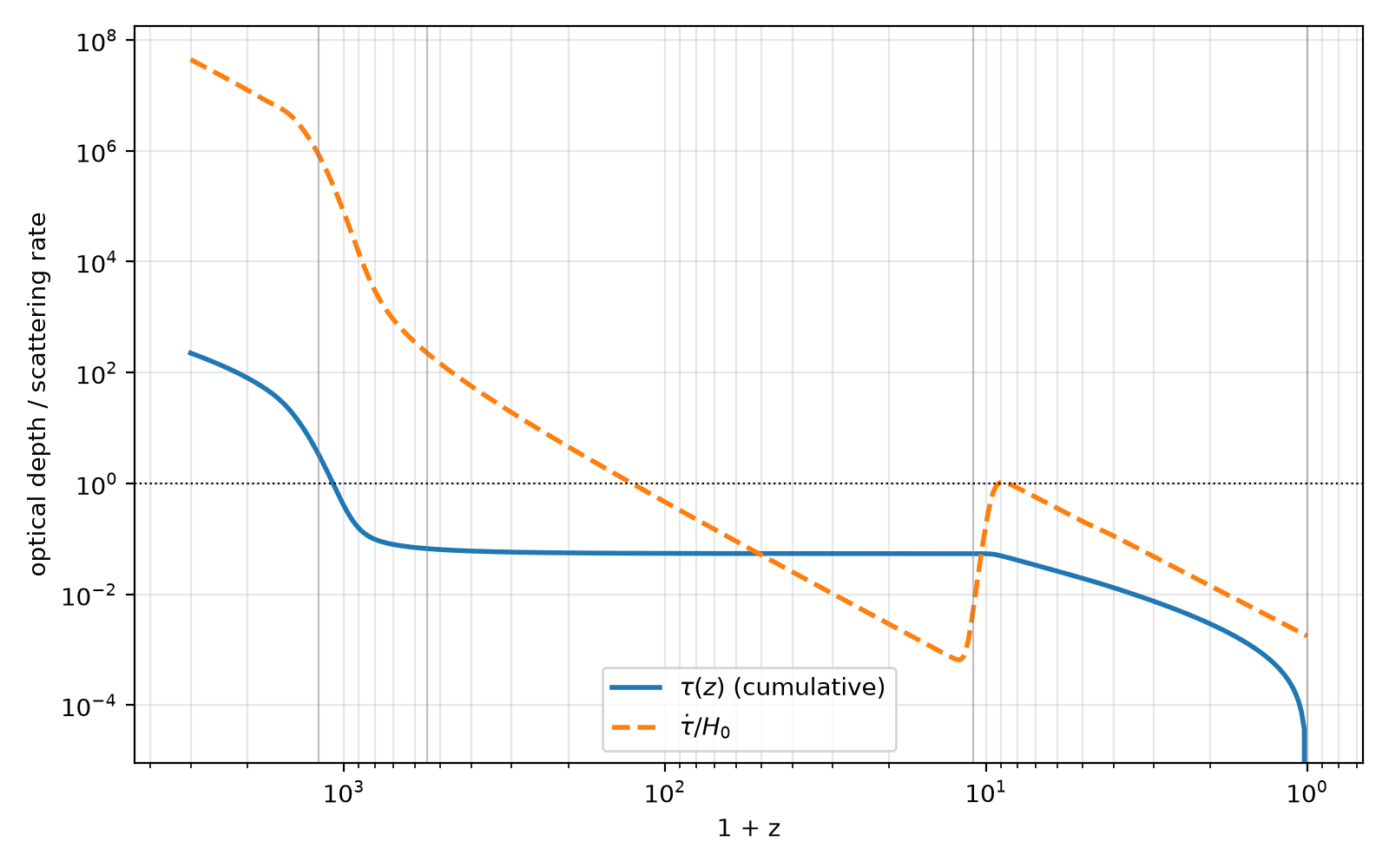}
	\caption{Variation of Thomson scattering rate (dashed orange, appearing in our equations \eqref{maint}) and cumulative optical depth (solid blue) against redshift in all Thurston geometries, explaining the damping of the maximum amplitude in signals from recombination up to today's epoch.}
	\label{tau}
\end{figure}

The evolution of the patterns are shown in figures \ref{R3} - \ref{Solv} in thermodynamic (CMB) units, evolution of geometric twist $\left(\theta, \phi, \psi \right)$ with redshift is shown in figure \ref{twist} and the evolution of optical depth $\left(\tau \right)$ is shown in figure \ref{tau} that aid in deciphering with redshift the shapes and amplitudes of signals so obtained in all the geometries. To completely specify the initial state and better isolate the geometries, the system is initialized with a purely real, initially unpolarized mode $\ell = 2$ aligned with the principal axes of the computational frame, in accordance with the initial seeding formalism developed in \S\ref{form}. The overall degree of polarization strongly depends on the ionization history through the differential scattering rate $\tau$. We note that the geometric relationship between the temperature and polarization patterns is strongly constrained by the \textit{geometrical structure} of our models.

We compute only the coherent part of the radiation field that arises from the geometry of the particular Thurston geometry under consideration. Any realistic cosmological model (one that produces galaxies and large-scale structures) must also have density inhomogeneities. Assuming these are of stochastic origin, they would add incoherent perturbations on top of the coherent ones produced by the background model.

%------------------------------------ CMB maps %------------------------------------
% ---- R3 ----
\newgeometry{margin=1cm, includefoot}
\restoregeometry

% ---- S3 ----
\newgeometry{margin=1cm, includefoot}
\restoregeometry

% ---- H3 ----
\newgeometry{margin=1cm, includefoot}
\restoregeometry

% ---- RS2 ----
\newgeometry{margin=1cm, includefoot}
\restoregeometry

% ---- RH2 ----
\newgeometry{margin=1cm, includefoot}
\restoregeometry

% ---- UH2 ----
\newgeometry{margin=1cm, includefoot}
\restoregeometry

% ---- Nil ----
\newgeometry{margin=1cm, includefoot}
\restoregeometry

% ---- Solv ----
\newgeometry{margin=1cm, includefoot}
\restoregeometry

%------------------------- Geometrical twist %-------------------------

\clearpage

%+++++++++++++++++++++++++++++++++++++++++++++++++++++++++++++++++++++++++++++++++++++++++++++++++++++++++++++++++++++++++++++++++++++++++++++++++++++++++++++++++++++++++++++
\section{Analysis \& conclusion} \label{five}
The goal of this work was to compute temperature and polarization patterns generated in each universe governed by one Thurston geometry at a time. We considered all the eight Thurston geometries described under geometrization conjecture, with all the them reducible to FLRW at the limit of vanishing curvature. We constructed an appropriate description of the radiation field in terms of spin-0 and spin-2 components representing the unpolarized and polarized parts, respectively, and integrated the equations for the radiation field numerically. With our analysis, we obtained rich temperature and polarization patterns and their evolution in all the Thurston geometries. The patterns are displayed for redshifts a little before recombination till today's epoch, and are dynamical enough for one to isolate individual geometries. In our results, the mean final temperature fluctuations are of $\mathcal{O} \left( \mu\text{K} \right)$, while polarization Stokes of $\mathcal{O} \left(10^{-4} \, \mu\text{K} \right)$, respectively, as generally expected for a CMB detector frequency of 150 GHz.

A visual inspection of the maps provide a qualitative analysis. In isotropic geometries, $\left(\mathbb{R}^3, S^3, \, \text{and} \, \mathbb{H}^3 \right)$, the patterns seem to \emph{mix} as we go from recombination to present redshift. In this regard, both temperature \& polarization patterns look identical across all these three isotropic geometries. On the other hand, in anisotropic geometries, the patterns of product (cylindrical) geometries look surprisingly distinguishable, in the sense that all  the patterns in $\mathbb{R} \times \mathbb{H}^2$ geometry seem to undergo a \emph{dipolar stretching}, while such prediction holds for temperature maps only for its closed counterpart $\mathbb{R} \times S^2$, even for opposite temperature hemispheres..  Notably, in $\mathbb{R} \times S^2$ geometry, the patterns do not change appreciably from dark ages to the present epoch, as compared to $\mathbb{R} \times \mathbb{H}^2$. Notefully, in these product geometries, the polarization rotation 2$\Delta \Psi$ stays at zero as being evaluated from eq. \eqref{polangle}, which is also demonstrated in figs. \ref{RS2_twist} \& \ref{RH2_twist}, respectively. Inherent curvature anisotropy geometries $\widetilde{U \left(\mathbb{H}^2 \right)}$, Nil \& Solv exhibit the richest evolutions of all. These geometries exhibit \emph{characteristic} spiraling and twisting patterns. The geometry $\widetilde{U \left(\mathbb{H}^2 \right)}$ exhibits a mix of rotation + distortion patterns, Nil geometry only the distortion ones, while Solv geometry exhibits pure rotational patterns. All the Thurston geometries, therefore, exhibit characteristic and mutually distinguishable set of patterns.

The basic point that emerges from this study is related to the physical origin of CMB polarization: radiation affected by Thomson scattering from an electron in a radiation field possessing a quadrupole anisotropy will inevitably be partially linearly polarized. In the context of standard cosmological models, the environment of different electrons varies owing to the presence of density inhomogeneities and a background of gravitational waves. These sources of variation are stochastic so the variations in the polarized component of the radiation field, though correlated with the temperature variations, are essentially incoherent. Present and future CMB experiments are largely aimed at improving the precision of polarization measurements in order to find evidence of a stochastic background of primordial gravitational waves. Techniques advocated to measure polarization pattern over the celestial sphere are therefore also generally tuned to detect signals of an incoherent nature. The presence of coherent signals in the CMB could be an indication of physics beyond the standard cosmological framework. This will hopefully lead us to better ways of searching for departures from the concordance cosmology using the next generation of datasets.

%+++++++++++++++++++++++++++++++++++++++++++++++++++++++++++++++++++++++++++++++++++++++++++++++++++++++++++++++++++++++++++++++++++++++++++++++++++++++++++++++++++++++++++++
\section{Acknowledgements}
The results in this paper have been derived using the HEALPix and healpy packages. We acknowledge the Python package \texttt{Healpy} \cite{zonca2019healpy} of the original HEALPix software \cite{gorski2005healpix, healpix_web} and the IISER Bhopal Gargi cluster.

%+++++++++++++++++++++++++++++++++++++++++++++++++++++++++++++++++++++++++++++++++++++++++++++++++++++++++++++++++++++++++++++++++++++++++++++++++++++++++++++++++++++++++++++}
\section{Code availability} \label{code}
All the codes along with the data files used in our work are publicly available at \url{https://github.com/Tanay80/Temp_polar_patterns_Thurston.git}.

%+++++++++++++++++++++++++++++++++++++++++++++++++++++++++++++++++++++++++++++++++++++++++++++++++++++++++++++++++++++++++++++++++++++++++++++++++++++++++++++++++++++++++++++}
\begin{appendices}
\section{Ricci rotation coefficients}
The exterior derivatives of the spatial basis 1-forms for each of the Thurston geometries are given by

\subsection{$\mathbb{R}^3$}
\begin{subequations}
\begin{gather}
	de^0 = 0 \\
	de^1 = H \left(e^0 \wedge e^1 \right) \\
	de^2 = H \left(e^0 \wedge e^2 \right) + \frac{1}{a \chi} \left(e^1 \wedge e^2 \right) \\
	de^3 = H \left(e^0 \wedge e^3 \right) + \frac{1}{a \chi} \left(e^1 \wedge e^3 \right) + \frac{1}{a \chi \tan \theta} \left(e^2 \wedge e^3 \right)
\end{gather}
\end{subequations}

\subsection{$S^3$}
\begin{subequations}
	\begin{gather}
		de^0 = 0 \\
		de^1 = H \left(e^0 \wedge e^1 \right) \\
		de^2 = H \left(e^0 \wedge e^2 \right) + \frac{\sqrt{k}}{a \tan \left(\chi \sqrt{k} \right)} \left(e^1 \wedge e^2 \right) \\
		de^3 = H \left(e^0 \wedge e^3 \right) + \frac{\sqrt{k}}{a \tan \left(\chi \sqrt{k} \right)} \left(e^1 \wedge e^3 \right) + \frac{\sqrt{k}}{a \sin \left(\chi \sqrt{k} \right) \tan \theta} \left(e^2 \wedge e^3 \right)
	\end{gather}
\end{subequations}

\subsection{$\mathbb{H}^3$}
\begin{subequations}
	\begin{gather}
		de^0 = 0 \\
		de^1 = H \left(e^0 \wedge e^1 \right) \\
		de^2 = H \left(e^0 \wedge e^2 \right) + \frac{\sqrt{-k}}{a \tanh \left(\chi \sqrt{-k} \right)} \left(e^1 \wedge e^2 \right) \\
		de^3 = H \left(e^0 \wedge e^3 \right) + \frac{\sqrt{-k}}{a \tanh \left(\chi \sqrt{-k} \right)} \left(e^1 \wedge e^3 \right) + \frac{\sqrt{-k}}{a \sinh \left(\chi \sqrt{-k} \right) \tan \theta} \left(e^2 \wedge e^3 \right)
	\end{gather}
\end{subequations}

\subsection{$\mathbb{R} \times S^2$}
\begin{subequations}
	\begin{gather}
		de^0 = 0 \\
		de^1 = H \left(e^0 \wedge e^1 \right) \\
		de^2 = H \left(e^0 \wedge e^2 \right) + \frac{\sqrt{k}}{a \tan \left(\chi \sqrt{k} \right)} \left(e^1 \wedge e^2 \right) \\
		de^3 = H \left(e^0 \wedge e^3 \right)
	\end{gather}
\end{subequations}

\subsection{$\mathbb{R} \times \mathbb{H}^2$}
\begin{subequations}
	\begin{gather}
		de^0 = 0 \\
		de^1 = H \left(e^0 \wedge e^1 \right) \\
		de^2 = H \left(e^0 \wedge e^2 \right) + \frac{\sqrt{-k}}{a \tanh \left(\chi \sqrt{-k} \right)} \left(e^1 \wedge e^2 \right) \\
		de^3 = H \left(e^0 \wedge e^3 \right)
	\end{gather}
\end{subequations}

\subsection{$\widetilde{U \left(\mathbb{H}^2 \right)}$}
\begin{subequations}
	\begin{gather}
		de^0 = 0 \\
		de^1 = H \left(e^0 \wedge e^1 \right) \\
		de^2 = H \left(e^0 \wedge e^2 \right) + \frac{\sqrt{-k} \tanh \left(x \sqrt{-k} \right)}{a} \left(e^1 \wedge e^2 \right) \\
		de^3 = H \left(e^0 \wedge e^3 \right) + \frac{\sqrt{-k}}{a} \left(e^1 \wedge e^2 \right)
	\end{gather}
\end{subequations}

\subsection{Nil}
\begin{subequations}
	\begin{gather}
		de^0 = 0 \\
		de^1 = H \left(e^0 \wedge e^1 \right) \\
		de^2 = H \left(e^0 \wedge e^2 \right) \\
		de^3 = H \left(e^0 \wedge e^3 \right) + \frac{\sqrt{-k}}{a} \left(e^1 \wedge e^2 \right)
	\end{gather}
\end{subequations}

\subsection{Solv}
\begin{subequations}
	\begin{gather}
		de^0 = 0 \\
		de^1 = H \left(e^0 \wedge e^1 \right) - \frac{\sqrt{-k}}{a} \left(e^1 \wedge e^3 \right)  \\
		de^2 = H \left(e^0 \wedge e^2 \right) + \frac{\sqrt{-k}}{a} \left(e^2 \wedge e^3 \right) \\
		de^3 = H \left(e^0 \wedge e^3 \right)
	\end{gather}
\end{subequations}
where $\left(e^a \wedge e^b \right) = -\left(e^b \wedge e^a \right) \, \forall \, a, b \in \left[0, 3 \right]$. The standard coordinate-free (invariant) formula for the exterior derivative `d' of a 1-form $\omega$ evaluated on two vector fields X and Y is given by
\begin{equation}
	d \omega \left(X, Y \right) = X \left(\omega \left(Y \right) \right) - Y \left(\omega \left(X \right) \right) - \omega \left(\left[X, Y \right] \right)
\end{equation}
Considering $\omega = e^a$, $X = e_b$ and $Y = e_c$ and $a \neq b \neq c \, \forall \, c \in \left[0, 3 \right]$ in general, we obtain
\begin{equation} \label{std}
	de^a \left(e_b, e_c \right) = -e^a \left(\left[e_b, e_c \right] \right)
\end{equation}
Substituting $\forall \, d \in \left[0, 3 \right]$
\begin{equation} 
	\left[e_b, e_c \right] = \gamma^d _{bc} \, e_d \, \text{\cite{sung2011temperature}}
\end{equation}
into equation \eqref{std}, we obtain
\begin{equation} \label{first_form}
	de^a \left(e_b, e_c \right) = -\gamma^a _{bc}
\end{equation}
Since we can write
\begin{equation} \label{second_form}
	de^a = \frac{1}{2} de^a \left(e_b, e_c \right) \left(e^b \wedge e^c \right)
\end{equation}
Then from equations \eqref{first_form} and \eqref{second_form}, we finally have
\begin{equation} \label{final_form}
	de^a = -\frac{1}{2} \gamma^a _{bc} \left(e^b \wedge e^c \right)
\end{equation}
Equation \eqref{final_form} is summed over indices b and c. Each component is given by
\begin{equation}
\begin{split}
	de^a &= -\frac{1}{2} \gamma^a _{bc} \left(e^b \wedge e^c \right) = -\frac{1}{2} \left[\gamma^a _{bc} \left(e^b \wedge e^c \right) + \gamma^a _{cb} \left(e^c \wedge e^b \right) \right] \\
	&= -\frac{1}{2} \left[\gamma^a _{bc} \left(e^b \wedge e^c \right) + \gamma^a _{bc} \left(e^b \wedge e^c \right) \right] \\
	&= -\gamma^a _{bc} \left(e^b \wedge e^c \right)
\end{split}
\end{equation}
where we used the property $\gamma^a _{bc} = -\gamma^a _{cb}$, just like $\left(e^a \wedge e^b \right) = -\left(e^b \wedge e^a \right)$. Using this result, we obtain the set of structure constants for each Thurston geometry as

\subsection{$\mathbb{R}^3$}
\begin{subequations}
\begin{gather}
	\gamma^1{} _{01} = \gamma^2{} _{02} = \gamma^3{} _{03} = -H \\
	\gamma^2{} _{12} = \gamma^3{} _{13} = -\frac{1}{a \chi} \\
	\gamma^3{} _{23} = -\frac{1}{a \chi \tan \theta}
\end{gather}
\end{subequations}

\subsection{$S^3$}
\begin{subequations}
\begin{gather}
	\gamma^1{} _{01} = \gamma^2{} _{02} = \gamma^3{} _{03} = -H \\
	\gamma^2{} _{12} = \gamma^3{} _{13} = -\frac{\sqrt{k}}{a \tan \left(\chi \sqrt{k} \right)} \\
	\gamma^3{} _{23} = -\frac{\sqrt{k}}{a \sin \left(\chi \sqrt{k} \right) \tan \theta}
\end{gather}
\end{subequations}

\subsection{$\mathbb{H}^3$}
\begin{subequations}
	\begin{gather}
		\gamma^1{} _{01} = \gamma^2{} _{02} = \gamma^3{} _{03} = -H \\
		\gamma^2{} _{12} = \gamma^3{} _{13} = -\frac{\sqrt{-k}}{a \tanh \left(\chi \sqrt{-k} \right)} \\
		\gamma^3{} _{23} = -\frac{\sqrt{-k}}{a \sinh \left(\chi \sqrt{-k} \right) \tan \theta}
	\end{gather}
\end{subequations}

\subsection{$\mathbb{R} \times S^2$}
\begin{subequations}
	\begin{gather}
		\gamma^1{} _{01} = \gamma^2{} _{02} = \gamma^3{} _{03} = -H \\
		\gamma^2{} _{12} = -\frac{\sqrt{k}}{a \tan \left(\chi \sqrt{k} \right)}
	\end{gather}
\end{subequations}

\subsection{$\mathbb{R} \times \mathbb{H}^2$}
\begin{subequations}
	\begin{gather}
		\gamma^1{} _{01} = \gamma^2{} _{02} = \gamma^3{} _{03} = -H \\
		\gamma^2{} _{12} = -\frac{\sqrt{-k}}{a \tanh \left(\chi \sqrt{-k} \right)}
	\end{gather}
\end{subequations}

\subsection{$\widetilde{U \left(\mathbb{H}^2 \right)}$}
\begin{subequations}
	\begin{gather}
		\gamma^1{} _{01} = \gamma^2{} _{02} = \gamma^3{} _{03} = -H \\
		\gamma^2{} _{12} = -\frac{\sqrt{-k} \tanh \left(x \sqrt{-k} \right)}{a} \\
		\gamma^3{} _{12} = -\frac{\sqrt{-k}}{a}
	\end{gather}
\end{subequations}

\subsection{Nil}
\begin{subequations}
	\begin{gather}
		\gamma^1{} _{01} = \gamma^2{} _{02} = \gamma^3{} _{03} = -H \\
		\gamma^3{} _{12} = -\frac{\sqrt{-k}}{a}
	\end{gather}
\end{subequations}

\subsection{Solv}
\begin{subequations}
	\begin{gather}
		\gamma^1{} _{01} = \gamma^2{} _{02} = \gamma^3{} _{03} = -H \\
		\gamma^1{} _{31} = -\gamma^2{} _{32} = -\frac{\sqrt{-k}}{a}
	\end{gather}
\end{subequations}
Finally, Ricci rotation coefficients for all the geometries are obtained using \cite{sung2011temperature} as
\begin{equation}
	\Gamma^d{} _{bc} = \eta^{ad} \Gamma_{abc} = \frac{1}{2} \eta^{ad} \left(\gamma_{abc} + \gamma_{cab} - \gamma_{bca} \right)
\end{equation}
\end{appendices}

%+++++++++++++++++++++++++++++++++++++++++++++++++++++++++++++++++++++++++++++++++++++++++++++++++++++++++++++++++++++++++++++++++++++++++++++++++++++++++++++++++++++++++++++}
\bibliography{cite}
\end{document}